\immediate\write18{makeindex -s nomencl.ist -o "\jobname.nls" "\jobname.nlo"} % For nomenclature

\documentclass[review,12pt]{elsarticle}

\usepackage{amssymb}
\usepackage{amsthm}
\usepackage{amsmath}
\usepackage{mathrsfs}
\usepackage{graphicx}
\usepackage{epstopdf}
\usepackage{float}
\usepackage{caption}
\usepackage[margin=2cm]{geometry}% TO PLAY WITH THE MARGINS
\usepackage{soul}
\usepackage{url}
\usepackage{color,soul} %to highlight text
\usepackage{booktabs}
\usepackage{color} %To highlight references in the text
\usepackage{subcaption}
\usepackage{bm}
\usepackage{bbm}
\usepackage{mathrsfs}
\usepackage{cleveref}
\usepackage{soul}
\usepackage{upgreek}
\usepackage{framed} % To put a box to the nomenclature
\usepackage{nomencl} %to add the nomenclature
\makenomenclature %to add the nomenclature
\biboptions{sort&compress}

\journal{International Journal of Hydrogen Energy}

\makeatletter
\def\@author#1{\g@addto@macro\elsauthors{\normalsize%
    \def\baselinestretch{1}%
    \upshape\authorsep#1\unskip\textsuperscript{%
      \ifx\@fnmark\@empty\else\unskip\sep\@fnmark\let\sep=,\fi
      \ifx\@corref\@empty\else\unskip\sep\@corref\let\sep=,\fi
      }%
    \def\authorsep{\unskip,\space}%
    \global\let\@fnmark\@empty
    \global\let\@corref\@empty  %% Added
    \global\let\sep\@empty}%
    \@eadauthor={#1}
}
\makeatother

\makeatletter
\def\thickhline{%
  \noalign{\ifnum0=`}\fi\hrule \@height \thickarrayrulewidth \futurelet
   \reserved@a\@xthickhline}
\def\@xthickhline{\ifx\reserved@a\thickhline
               \vskip\doublerulesep
               \vskip-\thickarrayrulewidth
             \fi
      \ifnum0=`{\fi}}
\makeatother

\newlength{\thickarrayrulewidth}
\begin{document}

\begin{frontmatter}



\title{TDS Simulator: A MATLAB App to model temperature-programmed hydrogen desorption}


\author{Enrique Garc\'{\i}a-Mac\'{\i}as\fnref{IC,UGR}}

\author{Zachary D. Harris\fnref{Pitt}}

\author{Emilio Mart\'{\i}nez-Pa\~neda\corref{cor1}\fnref{Ox}}
\ead{emilio.martinez-paneda@eng.ox.ac.uk}

\address[IC]{Department of Civil and Environmental Engineering, Imperial College London, London SW7 2AZ, UK}

\address[UGR]{Department of Structural Mechanics and Hydraulic Engineering, University of Granada, Granada 18002, Spain}

\address[Pitt]{Department of Mechanical Engineering and Materials Science, University of Pittsburgh, Pittsburgh, PA 15261, USA}

\address[Ox]{Department of Engineering Science, University of Oxford, Oxford OX1 3PJ, UK}

\cortext[cor1]{Corresponding author.}

\begin{abstract}
We present TDS Simulator, a new software tool aimed at modelling thermal desorption spectroscopy (TDS) experiments. TDS is a widely used technique for quantifying key characteristics of hydrogen-material interactions, such as diffusivity and trapping. However, interpreting the output of TDS experiments is non-trivial and requires appropriate post-processing tools. This work introduces the first software tool capable of simulating TDS curves for arbitrary choices of material parameters and hydrogen trap characteristics, using the primary hydrogen diffusion and trapping models (Oriani, McNabb-Foster). Moreover, TDS Simulator contains a specific functionality for loading experimental TDS data and conducting the inverse calibration of a selected transport model, providing automatic estimates of the density and binding energy of each hydrogen trap type in the material. In its first version, TDS Simulator is provided as a MATLAB App, which is made freely available to the community and provides a simple graphical user interface (GUI) to make use of TDS Simulator straightforward. As reported in the present manuscript, the outputs of TDS Simulator have been extensively validated against literature data. Demonstrations of automatic determination of trap characteristics from experimental data through the optimisation tool are also provided. The present work enables an efficient and straightforward characterisation of hydrogen-material characteristics relevant to multiple applications, from nuclear fusion to the development of hydrogen-compatible materials for the hydrogen economy. TDS Simulator can be downloaded from \url{https://mechmat.web.ox.ac.uk/codes} (To be uploaded immediately after the review process).\\

\end{abstract}

\begin{keyword}

Hydrogen \sep Diffusion \sep Thermal Desorption Spectroscopy  \sep Trapping \sep MATLAB



\end{keyword}

\end{frontmatter}


% Nomenclature
% The following order should be used within this table: Latin characters should appear first, arranged a, A, b, B etc.; then Greek characters, similarly arranged; sub/superscripts, abbreviations, special functions etc. usually come as a separate final group.
% makeindex EFM.nlo -s nomencl.ist -o EFM.nls
\begin{framed}
\nomenclature{$L$}{Thickness of the specimen [m]}
\nomenclature{$J$}{Hydrogen flux [mol$\,$m/s]}
\nomenclature{$C_L$}{Hydrogen concentration in lattice sites [mol/m$^3$]}
\nomenclature{$C_{T}^{(i)}$}{Hydrogen concentration in trapping sites of type $i$ [mol/m$^3$]}
\nomenclature{$C_L^0$}{Initial hydrogen concentration in lattice sites [mol/m$^3$]}
\nomenclature{$C_0$}{Total initial hydrogen concentration [mol/m$^3$]}
%\nomenclature{$\alpha$}{Number of atom sites per trap [-]}
\nomenclature{$\beta$}{Number of normal interstitial lattice sites per lattice atom [-]}
\nomenclature{$K^{(i)}$}{Equilibrium constant for the $i$th trap type [-]}
\nomenclature{$\theta_{T}^{(i)}$}{Occupancy of trap type $i$ [-]}
\nomenclature{$\theta_L$}{Occupancy of lattice sites [-]}
%\nomenclature{$\theta_{T,i}^0$}{Initial occupancy of trap type $i$ [-]}
\nomenclature{$\theta_L^0$}{Initial occupancy of lattice sites [-]}
\nomenclature{$D_0$}{Pre-exponential factor for lattice diffusion [m$^2$/s]}
%\nomenclature{$k_T^0$}{Pre-exponential constant for the capture rate [m$^3$/(mol$\cdot$s)]}
%\nomenclature{$p_T^0$}{Pre-exponential constant for the release rate [$s^{-1}$]}
\nomenclature{$k^{(i)}$}{Pre-exponential constant for the capture rate [-]}
\nomenclature{$p^{(i)}$}{Pre-exponential constant for the release rate [-]}
\nomenclature{$\nu_t^{(i)}$}{Vibration frequency of the hydrogen atom hopping from a lattice site to a trap [Hz]}
\nomenclature{$\nu_d^{(i)}$}{Vibration frequency of the hydrogen atom hopping from a trap to a lattice site [Hz]}
\nomenclature{$E_L$}{Activation energy for lattice diffusion [J/mol]}
\nomenclature{$N_{T}^{(i)}$}{Number of trap sites of type $i$ per unit volume [mol/m$^3$]}
\nomenclature{$N_L$}{Number of lattice sites per unit volume [mol/m$^3$]}
\nomenclature{$E_d^{(i)}$}{Activation energy for release (detrapping) of trap type $i$ [J/mol]}
\nomenclature{$E_t^{(i)}$}{Activation energy for capture (trapping) of trap type $i$ [J/mol]}
\nomenclature{$\Delta H^{(i)}$}{Trap binding enthalphy of trap type $i$ [J/mol]}
\nomenclature{$R$}{Universal gas constant [J/(mol$\cdot$K)]}
\nomenclature{$T$}{Temperature [K]}
\nomenclature{$T_{min},T_{max}$}{Minimum and maximum temperatures [K]}
\nomenclature{$\phi$}{Heating rate [K/s]}
\nomenclature{$t$}{Time [s]}
\nomenclature{$t_{\text{rest}}$}{Resting time [s]}
\printnomenclature
\end{framed}

\section{Introduction} \label{Introduction}
% *****************************************************************

% ----------------------------------------------------------------------------
% PARAGRAPH 1 - 
% Characterization of hydrogen-metal interactions (\textit{e.g.}, solubility, trapping, diffusivity, etc.) is critical for understanding hydrogen-induced degradation of structural metals. Provide a few short examples...
Hydrogen is a critical component of proposed decarbonization strategies due to its natural abundance, minimal projected environmental impact, and potential for decarbonizing traditionally hard-to-abate industries~\cite{pleshivtseva2023comprehensive,guo2024hydrogen,breyer2024role}. However, while it is clear that hydrogen offers significant promise as an energy carrier, a key impediment to the proliferation of a hydrogen-based economy is the propensity for hydrogen to degrade the mechanical properties of structural metals \cite{Gangloff2008}. This hydrogen embrittlement phenomenon has been responsible for several recent high-profile component failures, such as the well-publicized fracture of 32 anchor rods on the new eastern span of the San Francisco-Oakland Bay Bridge \cite{townsend2015hydrogen}. While many factors can influence the susceptibility of a material to hydrogen embrittlement, variations in hydrogen-metal interactions (\textit{e.g.}, hydrogen uptake, diffusion, and trapping) have a particularly strong effect. For example, the extent of degradation in alloy toughness or ductility has been shown to strongly depend on hydrogen concentration \cite{Moody1990,Gangloff2008,malheiros2022local}. Similarly, a strong correlation between the Stage II subcritical crack growth rate and hydrogen diffusivity has also been observed across a wide range of metallic materials \cite{Gangloff2003a}. As such, quantifying hydrogen-material interactions parameters can provide insights into embrittlement susceptibility; such insights are especially useful when developing new materials given that hydrogen diffusivity, uptake, and trapping behaviour are strongly influenced by alloy microstructure \cite{Chen2020,krom2000hydrogen,cupertino2023hydrogen}.\\

\begin{figure}[H]
\centering
\includegraphics[scale=1.2]{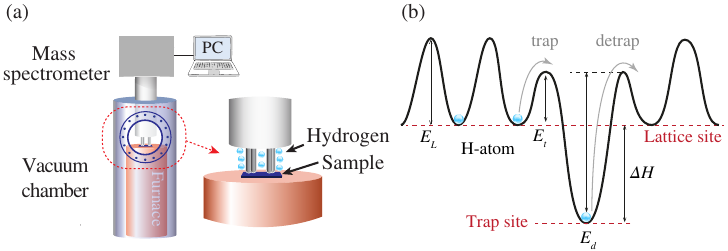}
\caption{Thermal desorption spectroscopy (TDS) to unravel hydrogen-material interactions; (a) Schematic illustration of a type of TDS apparatus, and (b) a schematic of the different energy levels involved in the diffusion of hydrogen in metals.}
\label{Scheme_TDS}
\end{figure}

% PARAGRAPH 2 - 
% There are several methods by which hydrogen-metal interactions can be assessed, including...
These dependencies have motivated the development of various experimental methods and techniques for assessing hydrogen-metal interactions \cite{li2020review, verbeken2012analysing}. For example, total hydrogen content can be determined via inert gas fusion, vacuum fusion, silicone oil, and laser thermal desorption ~\cite{brun2022machinery}, while techniques such as the Barnacle cell method enable quantification of the diffusible hydrogen content \cite{berman1988barnacle}. However, the two most widely adopted methods for evaluating hydrogen interactions with metallic microstructure are hydrogen permeation and thermal desorption spectroscopy (TDS) \cite{depover2021hydrogen,zafra2022comparison}. Briefly, permeation involves the generation and uptake of hydrogen on one side of a thin membrane (via gaseous hydrogen exposure or electrochemical hydrogen production) and then measuring the rate of hydrogen effusion on the other side via mass spectrometry, change in vacuum pressure, or oxidation current density \cite{van2023methodology}. The hydrogen flux versus time data are then evaluated using various theoretical models to determine the effective hydrogen diffusivity and diffusible hydrogen concentration for the employed environment/material combination. The benefits of electropermeation are the ease of implementation and straightforward analysis of generated data, but this method is prone to significant test-to-test variability \cite{turnbull1995factors,zafra2023relative,zafra2022comparison}. Conversely, TDS involves controlled out-gassing of a hydrogen pre-charged specimen as a function of temperature, with the hydrogen flux monitored via (e.g.) a high-resolution mass spectrometer (Fig. \ref{Scheme_TDS}). Analysis of the hydrogen flux versus temperature or time then enables the determination of hydrogen trapping characteristics (binding energies, site densities), effective diffusivity, and total hydrogen concentration \cite{verbeken2012analysing}. The benefits of TDS experiments are the ability to calculate most primary hydrogen-metal interaction parameters, but such experiments generally require an ultra-high vacuum environment and the use of uniformly pre-charged specimens, although ambient pressure systems are now commercially available (albeit with reduced hydrogen sensing resolution).

In addition to the need for specialized equipment, another challenge associated with TDS measurements is the interpretation of the desorption spectra \cite{cheng2013further,song2013theory,raina2017analysis,Raina2018}. Several theoretical frameworks have been proposed to relate TDS output to trapping characteristics, with the three most common ones being (1) McNabb and Foster \cite{McNabb1963}, (2) Oriani \cite{oriani1970diffusion}, and (3) Kissinger \cite{Kissinger1956,choo1982thermal}. McNabb and Foster's model provides a generalised treatment that explicitly includes hydrogen trapping and detrapping kinetics in the treatment of hydrogen diffusion \cite{mcnabb1963new,turnbull2015perspectives}. Oriani sought to simplify McNabb and Foster's framework by assuming a local equilibrium between trap sites and the lattice \cite{oriani1970diffusion}, which reduces the McNabb and Foster framework to a single equation \cite{sofronis1989numerical,thomas2002trap}. Lastly, Kissinger \cite{Kissinger1956} assumes a detrapping-dominated paradigm where diffusion is infinitely fast, which also simplifies the McNabb and Foster framework and enables straightforward determination of trap binding energies \cite{choo1982thermal}. Kissinger's method, also referred to as the Choo-Lee approach, is frequently used due to its simplicity but the assumptions employed result in a very narrow regime of validity. Specifically, since it assumes infinitely fast diffusion, it is only suitable for a small range of heating rates, sufficiently thin samples, and high-diffusivity materials \cite{Legrand2015,diaz2020influence,guterlrevisited}. Moreover, removing an effect of diffusion inherently precludes capturing the effect of specimen thickness, trapping densities or initial hydrogen concentration. For this reason, both Oriani and McNabb-Foster models are often of interest for analysing TDS spectra, but such approaches require the use of numerical tools which are not readily available across the hydrogen community. The present work aims to fill this need by providing the first generalized framework for analyzing TDS data, including both McNabb and Foster and Oriani-based analyses for an arbitrary number of traps, and the first standalone software package for conducting virtual TDS experiments. In this way, the present work contributes to ongoing efforts in the community aimed at providing software tools to facilitate an improved understanding of hydrogen-material interactions \cite{lototskyy2016new,charles2021numerical,delaporte2024festim}.\\

% PARAGRAPH 4 - 
% OBJECTIVE OF THIS WORK
In the remainder of this paper, we proceed to describe the characteristics of TDS Simulator, the new software tool that we have developed to automate the analysis of TDS data. TDS Simulator provides a GUI-based platform for creating synthetic TDS data and assessing experimental TDS spectra using the primary theories for hydrogen trapping and desorption. Critically, this MATLAB App enables the efficient determination of trapping parameters from experimental data using a deterministic variable inference algorithm. To establish appropriate context, we begin by providing a concise review of the relevant theories for modelling TDS data in Section \ref{Sec:Theory}. An overview of the software tool is then provided in Section \ref{Sec:Usage}. The robustness of the software package is demonstrated by validating against analytical and numerical data from the literature (Section \ref{Sec:Results1}), and its usage is exemplified by determining trapping characteristics from experimental TDS data (Section \ref{Sec:Results2}). Finally, the manuscript ends with concluding remarks in Section \ref{Sec:ConcludingRemarks}.

% *****************************************************************
\section{Theory: a generalised multi-trap framework for hydrogen transport}\label{Sec:Theory}
% *****************************************************************
Thermal desorption spectroscopy (TDS) experiments measure the hydrogen desorption rate from a sample of thickness $L$ that contains a given hydrogen concentration. Plate or disk-like samples are typically used. The specimen, uniformly precharged with a hydrogen concentration $C_L = C_L^0$ at $t = 0$ for all $x \in \left[-L/2,L/2\right]$, is located in a furnace (Fig.~\ref{Scheme_TDS} (a)). The hydrogen leaving the sample by diffusion, at the two parallel surfaces at $x=-L/2$ and $x=L/2$, is monitored by a mass spectrometer as the temperature rises from $T_0$ at a constant heating rate $\phi$. The problem is effectively one-dimensional, as $L$ is much smaller than any other sample dimension such that desorption takes place predominantly through the sample thickness. The hydrogen atoms in the metal can occupy normal interstitial lattice sites (NILS) as well as trapping sites, such as interfaces or dislocations. Thus, the total hydrogen concentration $C$ is the sum of lattice hydrogen concentration $C_L$ and trapped hydrogen concentration $C_T$.\\

\begin{figure}[H]
	\centering
	\includegraphics[scale=1.2]{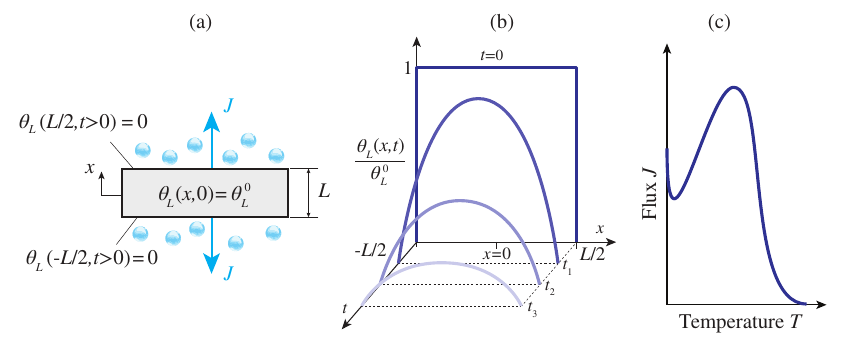}
	\caption{Hydrogen desorption in TDS experiments; (a) A schematic illustration of initial and boundary conditions in a TDS test; (b) Transient solution curves of the normalised lattice occupancy fraction $\theta_L/\theta^0_L$ at different times $t$ along the specimen's thickness; (c) A schematic of typical hydrogen desorption flux versus temperature curves obtained in a TDS test.}
	\label{Scheme_traps}	
\end{figure}

To model the trapping/detrapping kinetics of hydrogen atoms, the energy landscape for the diffusion of hydrogen in metals in Fig.~\ref{Scheme_TDS} (b) is commonly assumed. In this figure, the terms $E_t$ and $E_d$ denote the trapping and detrapping enthalpies, respectively. Specifically, $E_d$ is the activation energy required for hydrogen to move from a trap site to a lattice site, while $E_t$ is the activation energy for hydrogen to move from a lattice site to a trap site. On this basis, the trap binding energy is defined as $\Delta H=E_t-E_d$. The term $E_L$ stands for the diffusion activation energy. As the temperature rises, the lattice hydrogen concentration $C_L(x,t)$ evolves spatially and temporally as shown in Fig.~\ref{Scheme_traps} (b). In general, stronger traps release hydrogen at higher temperatures in a TDS experiment. As the temperature increases with the given rate, the trapped hydrogen is able to escape from the trap sites and diffuses out, resulting in a measured flux profile similar to that shown in Fig.~\ref{Scheme_traps} (c).

In this light, the evolution of the lattice concentration $C_L(x,t)$ of hydrogen atoms through a crystal lattice in the absence of traps is governed by Fickian diffusion over the lattice sites. Nevertheless, in the presence of traps, hydrogen diffusion is modified by both trapping and detrapping of hydrogen atoms. Mass conservation dictates that the rate of change of total concentration equals the net flux of diffusing hydrogen atoms. Assuming there are $n_t$ types of hydrogen traps, the governing equation of hydrogen diffusion can be written in the form of an extended one-dimensional Fick's second law as:  
\begin{equation}\label{Modif_Diff}
  \frac{\partial C_L}{\partial t} +  \sum_{i=1}^{n_t} \frac{\partial C_T^{(i)}}{\partial t}   = D_L \frac{\partial^2 C_L}{\partial x^2},
\end{equation}

\noindent where the superscript $i$ is used hereinafter to relate the corresponding quantity to the $i$-th trap, with a given trap binding energy $\Delta H^{(i)}$. Typically, each type of microstructural defect (dislocations, grain boundaries, carbides, etc.) is considered a distinct trap type and assigned a different $\Delta H^{(i)}$. But a higher level of detail can also be provided; e.g., different types of grain boundaries or dislocation regions can be considered different trap types, with different trap densities and binding energies.
The variable $D_L$ denotes the lattice diffusion coefficient, which is expressed in terms of the temperature $T = T_0+\phi t$, lattice activation energy $E_L$, diffusion pre-exponential factor $D_0$, and the universal gas constant $R$: $D_L=D_0 \exp(-E_L/RT)$. The heating rate $\phi$ is so slow compared to the rate of thermal diffusion that the TDS specimen is assumed to have a spatially uniform temperature $T(t)$. The hydrogen concentration in the lattice can be defined as $C_L=\theta_L N_L$, where $N_L$ denotes the number of interstitial sites per unit volume and $\theta_L$ is the lattice occupancy fraction ($0<\theta_L<1$). The former is typically estimated as $N_L = \beta N_A \rho_M / M_M$, where $\beta$ is the number of NILS per lattice atom ($\beta=6$ for bcc iron, $\beta=1$ for fcc iron), $N_A$ is Avogadro's number, $\rho_M$ is the material mass density, and $M_M$ is the molar mass. In the literature, traps are often deemed \emph{reversible} if they are weak traps (i.e.~low $|\Delta H|$), which readily release hydrogen, and \emph{irreversible} if they are strong traps (i.e.~high $|\Delta H|$). However, all trapping sites are reversible if a sufficiently wide range of time scales and temperatures is considered, and are therefore mathematically treated as such. Accordingly, the hydrogen concentration at reversible traps is given by $C_T=\theta_T  N_T$, where $N_T$ is the trap density and $\theta_T$ is the fractional occupancy of trap sites.\\

To solve the partial differential equation (PDE) in Eq.~(\ref{Modif_Diff}), numerical integration is required. To this aim, the initial and boundary conditions sketched in Fig.~\ref{Scheme_traps} (a) are considered. Initially, at time $t=0$, the specimen is at temperature $T_0$ with an initial uniform lattice occupancy $\theta_L=\theta_L^0$. Thereafter, the hydrogen lattice occupancy $\theta_L=0$ at the boundaries $x=\pm L/2$ is maintained. In other words, the boundary value problem has as initial condition $\theta_L(x,t=0) = \theta_L^0$, with $\theta_L^0$ being given by $\theta_L^0 = C_L^0 / N_L$, and the boundary condition reads $\theta_L(\pm L/2,t>0) = 0$. \\ 

Once solved, the flux of hydrogen atoms $J(t)$ diffusing out at the boundaries $x=\pm L/2$ can be estimated. Assuming $J(t)$ as the number of hydrogen atoms that exit the specimen per unit surface area, per unit time, the flux can be calculated from the concentration gradient at the surface of the sample ($x = \pm L/2$) as:
\begin{equation}\label{Flux_equation}
  J(\pm L/2,t) = -D_L\left.\frac{\partial C_L}{\partial x}\right|_{x=\pm L/2}.
\end{equation}

More often, experimental TDS spectra are shown in terms of the hydrogen desorption rate $\Delta C$, which can be estimated as
\begin{equation}\label{desprate}
\Delta C = \Delta C_L + \sum_{i=1}^{n_t} \Delta C_T^{(i)} =\frac{1}{L}\frac{\textrm{d}\int_{-L/2}^{L/2}C_L(x,t)\textrm{d}x}{\textrm{d}t} + \sum_{i=1}^{n_t} \frac{1}{L}\frac{\textrm{d}\int_{-L/2}^{L/2}C_T^{(i)}(x,t)\textrm{d}x}{\textrm{d}t}.
\end{equation}

To complete the definition of the governing model in Eq.~(\ref{Modif_Diff}), it remains to define the rate of trapped hydrogen concentration $\partial C_T^{(i)}/\partial t$ considering the kinetics of trapping and detrapping. In the following, three different approaches commonly adopted for interpreting TDS data are concisely overviewed.

% ----------------------------------------------------------------------------
\subsection{No Trapping}\label{Notraps}
% ----------------------------------------------------------------------------

Under the assumption of no trapping, Eq.~(\ref{Modif_Diff}) reduces to the classical Fick's Second Law. Assuming the diffusion coefficient $D_L$ to be temperature dependent, the diffusion equation can be solved in closed form by variable separation. Considering the aforementioned boundary and initial conditions, the lattice hydrogen concentration profile can then be obtained as \cite{crank1979mathematics,kirchheim2016bulk}:
\begin{equation}\label{notrap1}
    C_L(x,t) = \frac{4C_L^0}{\pi}\sum_{n=0}^{\infty} \frac{\left(-1\right)^n}{2n+1}\exp \left[ -\frac{\pi^2(2n+1)^2 D_{ft}(t)}{L^2}\right]\cos\left[ \frac{(2n+1)\pi x}{l} \right],
\end{equation}

\noindent where the term $D_{ft}$ is defined as:
\begin{equation}\label{notrap2}
    D_{ft}(t) = \int_0^t D_f \, \textrm{d}t = \frac{1}{\phi} \int_0^T D_L \, \textrm{d}T.
\end{equation}

% ----------------------------------------------------------------------------
\subsection{McNabb-Foster governing equations}\label{Sec_McNabb_Foster}
% ----------------------------------------------------------------------------

The general equilibrium equation proposed by McNabb and Foster reads~\cite{mcnabb1963new}: 
\begin{equation}\label{eq:RainaMcNabbFoster}
    \frac{\partial \theta_T^{(i)}}{\partial t}= \left[ \nu_t^{(i)} \exp \left( - \frac{E_t^{(i)}}{RT} \right) \theta_L \left( 1 - \theta_T^{(i)} \right)  - \nu_d^{(i)} \exp \left( - \frac{E_d^{(i)}}{RT} \right) \theta_T^{(i)} \left( 1- \theta_L \right) \right] \left( \frac{ N_L}{ N_L +  N_T^{(i)}} \right),
\end{equation}

\noindent where the terms $\nu_t^{(i)}$ and $\nu_d^{(i)}$ respectively denote the vibration frequency of the hydrogen atom hopping from a lattice site to a trap and from a trap to a lattice site. It is commonly assumed that $\nu_t^{(i)}=\nu_d^{(i)}=\nu$ (usually taken to be equal to the Debye frequency, i.e.~$\nu=10^{13}$ Hz). \\

Assuming that traps represent local defects in the lattice, the number of trap sites is commonly assumed to be much smaller than the number of lattice sites, that is $N_T << N_L$; i.e., $ N_L/ ( N_L +  N_T^{(i)} ) \approx 1$. On this basis, the equilibrium equation in Eq.~(\ref{eq:RainaMcNabbFoster}) reduces to:
\begin{equation}\label{eq:RainaMcNabbFoster_b}
    \frac{\partial \theta_T^{(i)}}{\partial t}= \left[ k^{(i)} \theta_L \left( 1 - \theta_T^{(i)} \right)  - p^{(i)} \theta_T^{(i)}\left( 1 - \theta_L \right)  \right],
\end{equation}

\noindent with
\begin{equation}\label{eq:RainaMcNabbFoster_c}
k^{(i)} = \nu_t^{(i)} \exp \left( - \frac{E_t^{(i)}}{RT} \right), \quad p^{(i)} = \nu_d^{(i)} \exp \left( - \frac{E_d^{(i)}}{RT} \right).
\end{equation}

Note that this equilibrium equation implies solving another PDE (per trap type) along with the extended diffusion equation in Eq.~(\ref{Modif_Diff}). In steels, it is commonly assumed that $\theta_L << 1$ ($\therefore 1-\theta_L \approx 1$), which may further reduce Eq.~(\ref{eq:RainaMcNabbFoster_b}); for the sake of generality, this simplification is not adopted here. Finally, one should note that the equilibrium equation (\ref{eq:RainaMcNabbFoster_b}) requires the definition of the initial values of the trap occupancies $\theta_T^{(i)}(x,t=0)=\theta_{T,0}^{(i)}$. Here, Oriani's equilibrium, discussed below, is adopted to define $\theta_{T,0}^{(i)}$.

% ----------------------------------------------------------------------------
\subsection{Oriani's model}\label{Oriani}
% ----------------------------------------------------------------------------

Oriani~\cite{oriani1970diffusion} assumed that trap kinetics occur on a much smaller time scale than diffusion of hydrogen through the lattice. Such an assumption implies that the chemical potentials of the hydrogen in lattice sites and in trap sites are equal, which results in the time-derivative in Eq.~(\ref{eq:RainaMcNabbFoster_b}) being zero~\cite{krom2000hydrogen}. Assuming $\nu_t^{(i)}=\nu_d^{(i)}=\nu$, this leads to:
\begin{equation}\label{Oriani_a}
\frac{\theta_T^{(i)}}{1-\theta_T^{(i)}} = \frac{\theta_L}{1-\theta_L}K_T^{(i)},
\end{equation}

\noindent with the equilibrium constant $K_T^{(i)}$ given by:
\begin{equation}\label{Oriani_b}
K_T^{(i)} =\exp \left(\frac{-\Delta H^{(i)} } {RT}\right).
\end{equation}

\noindent Note that Oriani's equilibrium (\ref{Oriani_a}) can be also derived by considering $\nu \rightarrow \infty$ in Eq.~(\ref{eq:RainaMcNabbFoster_b}) at finite $\partial \theta_T^{(i)}/\partial t$. \\

Introducing Eq.~(\ref{Oriani_a}) into the diffusion model in Eq.~(\ref{Modif_Diff}), one can write:
\begin{equation}\label{Oriani_c}
\frac{\partial \theta_L}{\partial t}\left\{1+\sum_{i=1}^{n_t} \frac{N_T^{(i)} K_T^{(i)}}{ N_L \left[1+\left(K_T^{(i)}-1\right)\theta_L \right]^2}\right\}+\sum_{i=1}^{n_t}\frac{N_T^{(i)} K_T^{(i)} \Delta H^{(i)} \phi(\theta_L-\theta_L^2)}{ N_L RT^2 \left[1+(K_T^{(i)}-1)\theta_L \right]^2} = D_L  \frac{\partial^2 \theta_L}{\partial x^2},
\end{equation}

\noindent which reduces the problem to one single second-order PDE that exclusively depends on $\theta_L$. Oriani's model can also be derived from generalised thermodynamic potentials, as done by Svoboda and Fischer \cite{svoboda2012modelling}. It is also worth noting that Kirchheim \cite{kirchheim2016bulk} derived approximate analytical expressions based on Oriani and showed how these could result in a Kissinger-like equation that incorporates diffusion information.

% *****************************************************************
\section{TDS Simulator}\label{Sec:Usage}
% *****************************************************************

% ----------------------------------------------------------------------------
\subsection{Software basics and architecture}
\begin{figure}[H]
	\centering
	\includegraphics[width=0.99\textwidth]{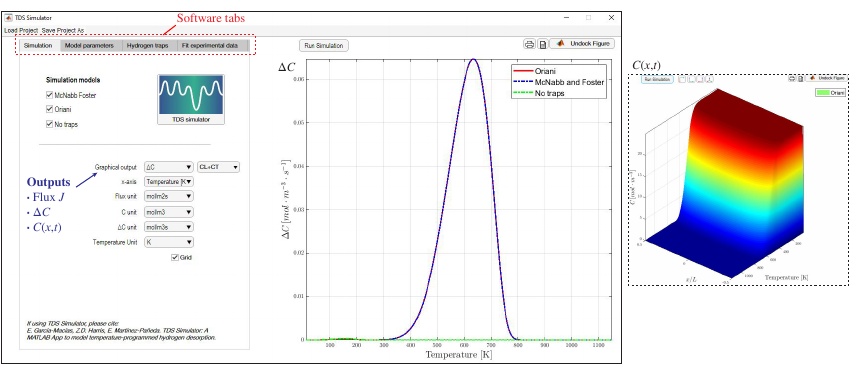}
	\caption{TDS Simulator: Main graphical user interface (GUI) of the software.}
	\label{Software_graph_1}	
\end{figure}

% ----------------------------------------------------------------------------
The previous formulation has been coded in MATLAB environment as a compact \texttt{AppDesigner} graphical user interface (GUI). The software is accessible as a standard MATLAB toolbox and provided as MATLAB App Installer File \texttt{TDS\_Simulator.mlappinstall}. The App can be downloaded from \url{https://mechmat.web.ox.ac.uk/codes} (To be uploaded immediately after the review process). One can readily install the TDS Simulator package by double-clicking the App Installer file or through MATLAB's \texttt{Install App} button within the \texttt{Apps} tab. Upon clicking on the TDS Simulator App icon, the main graphical user interface (GUI) shows up, as depicted in Fig.~\ref{Software_graph_1}. Each of the options available within the different tabs provided is discussed below.

In regards to the software architecture, the coupled partial differential equations (PDEs) of the McNabb-Foster's model, Eqs. (\ref{Modif_Diff}) and (\ref{eq:RainaMcNabbFoster_b}), are solved using finite differences and the \texttt{ode15s} solver in MATLAB. The \texttt{ode15s} solver is a variable-order solver based on numerical differentiation formulas (NDFs), which is especially well-suited for the solution of stiff problems. On the other hand, Oriani's model, Eq. (\ref{Oriani_c}), is solved by using the PDE solver \texttt{pdepe} in MATLAB. The \texttt{pdepe} solver is based on the method of lines which converts the given PDE into a system of initial value problems. In this method, the spatial derivatives are replaced with algebraic approximations and the remaining time derivatives are solved as a system of ordinary differential equations. An automatic time-stepping routine in the \texttt{pdepe} solver ensures that temporal convergence is achieved in each solution step. It should also be noted that Oriani-based predictions are obtained using a non-dimensional form of the governing equation. All the relevant equations of this work (Oriani, McNabb-Foster) are provided in non-dimensional form in \ref{App:NonDimensional} for the sake of generality, but the present implementation only handles Oriani's model in non-dimensional form. The App also includes an optimization module to automatically determine trap characteristics from experimental data. The characteristics of this module, which requires access to MATLAB's \texttt{Optimization} toolbox, are discussed below. In addition, the App includes the possibility of saving (and loading) projects through the \texttt{Save Project As} (and \texttt{Load Project}) tabs. This enables saving the set of parameters and results as a \texttt{.mat} file, for subsequent use or exchange among users. Finally, as visible in Fig.~\ref{Software_graph_1}, there is a \texttt{Run Simulation} button that starts the simulation and promptly provides its output on the graph below (typical simulation times are on the order of seconds). The output of the simulation can be saved as a Matlab figure (\texttt{.fig} extension, printer-like icon on the top left) or as a text file (\texttt{.txt} extension, document-like icon on the top left). In addition, one can also undock the figure.

% ----------------------------------------------------------------------------
\subsection{Elements of the App}
\begin{figure}[H]
	\centering
	\includegraphics[width=1.00\textwidth]{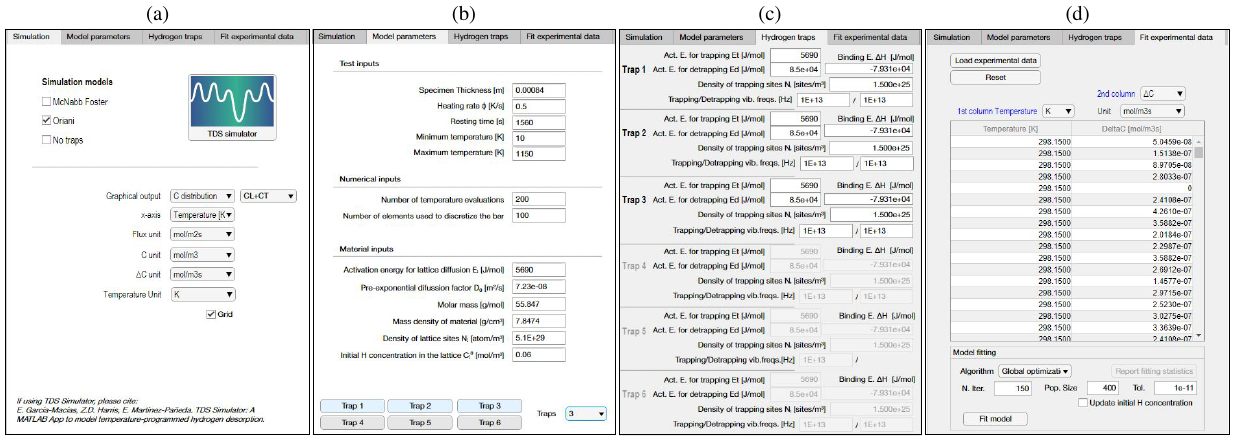}
	\caption{TDS Simulator: \texttt{Simulation} tab (a), \texttt{Model parameters} tab (b), \texttt{Hydrogen traps} tab (c), \texttt{Fit experimental data} tab (d).}
	\label{Software_tabs}	
\end{figure}

% ----------------------------------------------------------------------------
As shown in Fig. \ref{Software_tabs}, the TDS Simulator App includes four main tabs, aimed at: (i) establishing the characteristics of the simulation (\texttt{Simulation} tab), (ii) defining the material, numerical, and test parameters (\texttt{Model parameters} tab), (iii) describing the trap characteristics (\texttt{Hydrogen traps} tab), and (iv) fitting experimental data (\texttt{Fit experimental data} tab). Each of these is described in detail below. 

\subsubsection{\texttt{Simulation} tab}

As shown in Fig. \ref{Software_tabs} (a), the \texttt{Simulation} tab is aimed at defining the choice of hydrogen transport theory and specifying output characteristics (including units). First, the user must select the hydrogen trapping model to be employed, among the three described in Sections~\ref{Notraps} to \ref{Oriani}; i.e., no traps (lattice only), Oriani, and McNabb-Foster. It should be noted that the App allows selecting multiple choices simultaneously, to facilitate the comparison of the output of the different models. Additionally, the user can select the graphical options of the analysis for both the vertical and horizontal axes of the graph. For the horizontal axis, a typical output is the temperature $T$, but time $t$ is also an option. For the vertical axis, the user can pick between showing the flux $J$, as per Eq. (\ref{Flux_equation}), the hydrogen desorption rates $\Delta C$, as per Eq. (\ref{desprate}), or the hydrogen distribution as a function of space and time (temperature). The last option, shown in the inset of Fig.~\ref{Software_graph_1}, allows the user to visualise the distribution of lattice ($C_L(x,t)$), trapped ($C_T(x,t)$), or total ($C(x,t)$) hydrogen concentrations within the sample, as a function of temperature, using a 3D plot. For the case of the desorption rate, the total hydrogen desorption rate ($\Delta C$, \texttt{CL+CT}) is often the quantity of interest, as it is the output of the experiment; however, as discussed below in one of the examples, plotting the lattice ($\Delta C_L$, \texttt{CL}) and trapped ($\Delta C_T$, \texttt{CT}) desorption rates is also useful to gain insight into hydrogen partitioning. For each of these output quantities ($J$, $\Delta C$, $C(x,t)$), the user is asked to pick among the most typically used units. For the flux, these are mol/(m$^2$$\cdot$s), mol/(cm$^2$$\cdot$s) and wt ppm$\cdot$m/s, while  mol/m$^3$, mol/cm$^3$ and wt ppm are available for the hydrogen content, and mol/(m$^3$$\cdot$s), mol/(cm$^3$$\cdot$s) and wt ppm/s can be used for the desorption rate. Finally, the user also has the flexibility of including a grid in the graphical output if desired.

\subsubsection{\texttt{Model parameters} tab}

The inputs to the TDS experiment are provided in the \texttt{Model parameters} tab, shown in Fig.~\ref{Software_tabs} (b). There are essentially three categories of inputs: test inputs, numerical inputs, and material parameters. Test inputs include the sample thickness ($L$, in meters), the heating rate ($\phi$, in K/s), the resting time ($t_{\text{rest}}$, in s), and the minimum and maximum temperatures ($T_{\text{min}}$ and $T_{\text{max}}$, respectively, in K). The resting time refers to the time between the end of the hydrogen charging process and the beginning of the TDS test. During this resting period, the sample is held at room temperature for times that typically vary between 5 and 45 min., depending on the experimental approach adopted. Accordingly, the temperature variation is expressed as,
\begin{equation}\label{trest}
T=T_0+\phi \left\langle t-t_{\text{rest}}\right\rangle,
\end{equation}

\noindent where $\left\langle \cdot \right\rangle$ stands for Macaulay brackets (i.e.~$\left\langle x \right\rangle = x$ if $x \geq 0$, otherwise $\left\langle x \right\rangle = 0$).\\

Two numerical parameters are defined, the number of temperature evaluations (temperature discretisation) and the number of elements used to discretise the bar (sample thickness). The former mainly governs the number of data points shown as an output of a simulation; the default value of 200 has been shown to provide a sufficiently smooth representation in the case studies tested, but users can increase this number if needed (or reduce it, to achieve small computational gains). The number of elements used to discretise the bar has to be sufficiently large to result in a converged solution. The default number of 100 has proven to be sufficiently large to deliver an accurate result in all the case studies considered here and is therefore likely to be suitable for most if not all problems. However, the user can readily conduct a sensitivity study. This might be important before running optimisation analyses that may involve thousands of simulations, where calculation times are very sensitive to the number of elements used and a converged result might be achieved with a coarser mesh.\\

The last category of inputs in the \texttt{Model parameters} tab corresponds to the material parameters. These include the activation energy for lattice diffusion $E_L$ (in J/mol), the pre-exponential diffusion factor $D_0$ (in m$^2$/s), the molar mass $M_M$ (in g/mol), the mass density $\rho_M$ (in g/cm$^3$), and the density of lattice sites $N_L$ (in atomic sites/m$^3$). The molar mass $M_M$ and the mass density $\rho_M$ are only used for unit conversion purposes. These quantities are also related to the density of lattice sites $N_L$, e.g., for bcc iron
\begin{equation}\label{Eq:NLestimate}
    N_L = \frac{\beta N_A \rho_M}{M_M} = \frac{6  \,[\text{sites}/\text{at}] \cdot 6.022 \times 10^{23}  \, [\text{at}/\text{mol}] \cdot 7847.4   \,[\text{kg}/\text{m}^3]}{0.0558  \, [\text{kg}/\text{mol}]} = 5.1 \times 10^{29} \, [\text{sites}/\text{m}^3]
\end{equation}

But, for the sake of flexibility, $N_L$ is not automatically estimated and the user is instead asked to provide its magnitude as an input. It is worth emphasising that the formulation adopted here integrates into $N_L$ the density of the host metal lattice (solvent atoms per unit volume) and $\beta$, the number of lattice sites per atom; i.e., here $N_L$ is equivalent to what some papers in the literature denote as $\beta N_L$. As a result, the magnitude of $N_L$ in bcc iron is $5.1 \times 10^{29}$ sites/m$^3$, as per Eq. (\ref{Eq:NLestimate}), and not $8.46 \times 10^{28}$ at/m$^3$. The same applies to the trap density, where $N_T$ integrates the number of traps per unit volume and the number of atom sites per trap, often referred to as $\alpha$ in the literature (i.e., here $\alpha \equiv 1$). Some typical values of the material inputs required in TDS Simulator are provided in Table \ref{Tab:properties}. Since the properties provided refer to lattice characteristics, the materials included in Table \ref{Tab:properties} are nominally representative of that broader alloy class. For example, the bcc iron data is suitable for ferritic steels, such as pipeline steels, but is also a reasonable approximation for other carbon and low-alloy steels, such as bainitic and martensitic steels.

\begin{table}[H]
 \centering
  \caption{Typical values of the relevant material parameters, representative of a wide range of metals and alloys. Data taken from the literature \cite{Hirth1980,san2012technical,Raina2018,Gangloff2012,Bechtle2009,Young1998}. The estimate for the initial lattice content $C_L^0$ is based on Sievert's law, Eq. (\ref{Eq:Sievert}), for room temperature and 30 MPa H$_2$ exposure conditions.}
  \label{Tab:properties}
\resizebox{\textwidth}{!}{\begin{tabular}[t]{cccccccc}
\toprule
\text{Metal/alloy family} & $E_L$ [J/mol] & $D_0$ [m$^{2}$/s] & $M_M$ [g/mol] & $\rho_M$ [g/cm$^3$] &  $N_L$ [atom/m$^3$] & $S$ [mol/(m$^3$$\sqrt{\text{MPa}})$] & $C_L^0$ [mol/m$^3$]\\
\midrule
Bcc iron & 5690 & 7.23$\times 10^{-8}$  & 55.847 & 7.847 & 5.1$\times 10^{29}$ &  
0.011 & 0.06\\
Nickel & 40200 & 6.44$\times10^{-7}$  & 58.693 & 8.9 & 9.1$\times 10^{28}$ & 2.245  & 19.30 \\
Aluminium & 16200 & 1.8$\times 10^{-8}$ & 26.982 & 2.7 & 1.2$\times 10^{29}$ & 2.5$\times 10^{-6}$ &  1.3$\times 10^{-5}$ \\
Austenitic steel & 53600 & 6.2$\times 10^{-7}$ & 55.847 & 7.847 & 8.5$\times 10^{28}$ & 15.940 & 87.31 \\
\bottomrule
\end{tabular}}
\end{table}

It must be noted that Table \ref{Tab:properties} includes the initial hydrogen concentration in the lattice, $C_L^0$, which is related to both the material under consideration (solubility) and the test conditions (hydrogen charging environment). Its quantification is relatively straightforward for gaseous hydrogen (H$_2$) charging conditions, using Sievert's law \cite{san2012technical}:
\begin{equation}\label{Eq:Sievert}
    C_L = S \sqrt{p_{\text{H}_2}}
\end{equation}
\noindent where $S$ is the material solubility and $p_{\text{H}_2}$ is the H$_2$ pressure. However, its quantification for aqueous electrolyte environments is significantly more complex \cite{CS2022,cupertino2024hydrogen}. Nevertheless, as discussed below, predictions are not so sensitive to this input and therefore providing an initial estimate for $C_L^0$ within the right order of magnitude is sufficient to obtain accurate results over a wide range of environments. To facilitate this, Table \ref{Tab:properties} includes the lattice hydrogen concentration that corresponds to room temperature and 30 MPa H$_2$ exposure conditions. \\

Finally, the user must specify, at the bottom of the \texttt{Model parameters} tab, the number of trap types to be expected for this material ($n_t$).

\subsubsection{\texttt{Hydrogen traps} tab}

The \texttt{Hydrogen traps} tab, shown in Fig.~\ref{Software_tabs} (c) compiles the characteristics of up to 6 different trap types, with the relevant information requested depending on the hydrogen transport model selected in the \texttt{Simulation} tab. For the choice of Oriani, the inputs are the trap density $N_T$ (in sites/m$^3$) and the trap binding energy $\Delta H$ (in J/mol), with the latter being a negative quantity. For the McNabb-Foster model, in addition to the trap density $N_T$, the user can introduce the activation energy for trapping ($E_t$) and detrapping ($E_d$), with the trap binding energy being automatically estimated from these, $\Delta H = E_t - E_d$. By default, as is common in the literature, $E_t$ is taken to be the same as the lattice activation energy $E_L$. Therefore, if one changes the value of $\Delta H$, $E_t$ remains constant and $E_d$ is varied accordingly. In addition, the user can input the trapping and detrapping vibration frequencies, $\nu_t$ and $\nu_b$, respectively. These are set to the Debye frequency $\nu=\nu_t=\nu_d=10^{13}$ Hz by default. For both transport models, trap inputs must be introduced for every trap type, with the number of trap types being previously selected in the \texttt{Model parameters} tab. As described below, when the fitting algorithm is employed, these trapping variables are varied until a good fit with the experimental data is attained. 

\subsubsection{\texttt{Fit experimental data} tab}
\label{Sec:OptimizationProc}

The \texttt{Fit experimental data} tab enables uploading experimental data and inferring the trapping parameters that best describe these data by using an optimisation-based algorithm. TDS Simulator reads \texttt{.txt} files with two columns, the first one of which should be the temperature, while the second one can be desorption rate (\texttt{DeltaC}) or flux (\texttt{Flux}). As shown in Fig. \ref{Software_tabs}(d), the user must specify the variable of the second column ($\Delta C$ or $J$) and the units of both columns. Once uploaded, the set of experimental data points is automatically plotted in the graph, facilitating comparison with numerical outputs. The possibility to see both experimental and numerical TDS spectra in the same graph facilitates a manual, trial-and-error fitting procedure by the user if desired. However, as described below and shown in the bottom part of Fig. \ref{Software_tabs}(d), an automatic procedure is also included. 

\begin{figure}[H]
	\centering
	\includegraphics[width=\textwidth]{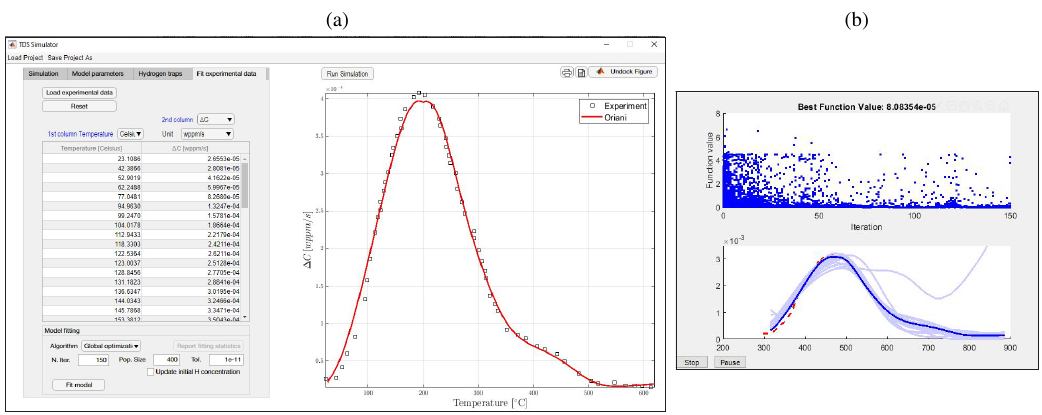}
	\caption{Using TDS Simulator to determine trapping characteristics by the automatic fitting of experimental data: (a) screenshot of the module used for the introduction of experimental data, and (b) graphical output of the parameter inference process.}
	\label{Software_graph_2}	
\end{figure}

TDS Simulator includes a specific module for loading experimental TDS data and inferring trapping parameters from this data, as shown in Fig.~\ref{Software_graph_2}. Specifically, the software includes a deterministic parameter inference algorithm based on the particle swarm algorithm (PSO) implemented in MATLAB, which requires access to the \texttt{Optimization Toolbox}. The PSO is a bio-inspired evolutionary algorithm, particularly well-suited for the solution of non-convex optimization problems. This algorithm searches the space of an objective function by adjusting the trajectories of individual solutions, called particles, as the piecewise paths formed by positional vectors in a quasi-stochastic manner~\cite{wang2018particle}. In TDS Simulator, the objective function is simply defined as the mean squared error between the experimental hydrogen desorption curve and the predictions of the selected model, that is:
\begin{equation}\label{objective_fun}
F(\boldsymbol{\uptheta}) = \sqrt{\mathbb{E}\left[\left(\Delta C_{exp}-\Delta C_{num}(\boldsymbol{\uptheta})\right)^2\right]},
\end{equation}

\noindent with $\mathbb{E}$ denoting the expectation operator. The minimization of $F(\boldsymbol{\uptheta})$ results in the following constrained non-linear optimization problem:
\begin{equation}\label{optimization}
\boldsymbol{\uptheta} = \textrm{min}_{\boldsymbol{\uptheta} \, \in \, \mathbb{D}} \, F(\boldsymbol{\uptheta}),
\end{equation}

\noindent with the vector $\boldsymbol{\uptheta}$ containing the trap parameters to be inferred and constrained within a certain physically meaningful range $\mathbb{D}$ described below. For simplicity, the trapping energy is fixed to the lattice activation energy, i.e.~$E_t^{(i)} = E_L$ \cite{Krom2000} and $\nu=\nu^t=\nu^d$ is assumed, limiting the software to infer the values of the binding energy $\Delta H^{(i)}$ and the density $N_T^{(i)}$ of the defined traps, for both Oriani and McNabb-Foster models.\\ 

Two different optimization approaches are available in TDS Simulator, including a global and a local search approach. These two approaches are equivalent and only differ in the allowed range of variation in $\mathbb{D}$. The global approach restraints the optimization problem in Eq.~(\ref{optimization}) to a broad variation range, that is $\mathbb{D}:=\left\{-150 \, \textrm{kJ/mol} \leq \Delta H^{(i)} \leq -15 \, \textrm{kJ/mol}; N_L \cdot 10^{-8} \leq N_T^{(i)} \leq N_L \cdot 10^{-1}\right\}$. On the other hand, the local search approach forces the optimizer to look for solutions in a neighbourhood of values that are within 80-120\% of the nominal values manually introduced by the user in the software (Fig.~\ref{Software_graph_2} (c)). As the optimization progresses, the software displays a graph showing the convergence of the solutions and the optimal flux solutions, see Fig.~\ref{Software_graph_2} (b). While these two approaches have proven to work well over the datasets considered so far, there are numerous other optimization algorithms that can be employed to enhance this fitting procedure. To facilitate the inverse calibration, the user is recommended to use the options available in the \texttt{Simulation} tab (Fig.~\ref{Software_tabs} (a)) for displaying hydrogen desorption, including total desorption $\Delta C$, lattice desorption $\Delta C_L$, and trap desorption $\Delta C_T^{(i)}$. These results can provide insight into the contribution of hydrogen traps and the lattice to the total desorption, helping to define the number of traps during calibration. Additionally, if the calibration results are not satisfactory, possibly due to the lack of convexity in the optimization problem, the user is advised to increase the number of particles (population size) and decrease the optimization tolerance. \\

It is important to note that TDS Simulator allows adjusting the initial hydrogen concentration as the optimization progresses. Note that, when an experimental $\Delta C$ versus $T$ curve is available, the area under the curve gives the total hydrogen concentration, $C_{exp}$. Since there is a finite amount of hydrogen in the metallic sample, then we have $C_L^0+\sum_{i=1}^{nt}\left(C_T^0\right)^{(i)}=C_{exp}$. Considering Oriani's equilibrium in Eq.~(\ref{Oriani_a}):
\begin{equation}\label{equilibrium_CT0}
\left(C_T^0\right)^{(i)} =  N_T^{(i)}\frac{K_T^{(i)}C_L^0}{ N_L+C_L^0\left(K_T^{(i)}-1\right)},
\end{equation}

\noindent and, therefore, the mass conservation law can be rewritten as:
\begin{equation}\label{Hydro_cons}
C_L^0+\sum_{i=1}^{nt}  N_T^{(i)}\frac{K_T^{(i)}C_L^0}{ N_L+C_L^0\left(K_T^{(i)}-1\right)}=C_{exp},
\end{equation}

Hence, when fitting experimental curves, TDS Simulator can follow the following iterative algorithm: (i) Pick a value for $N_T^{(i)}$ and $\Delta H_T^{(i)}$, giving $K_T^{(i)}$ according to Eq.~(\ref{Oriani_b}); (ii)  Estimate $C_L^0$ from Eq.~(\ref{Hydro_cons}); (iii) Obtain the resulting TDS curve and evaluate the objective function in Eq.~(\ref{optimization}), returning to (i) until the user-defined tolerance is met or the maximum number of iterations is reached. In this way, the predicted trapping values are consistent and $C_L^0$ no longer becomes a user input but is instead determined from the data. Nonetheless, note that this option may compromise the convergence of the inverse calibration and, therefore, it should be applied with caution. In addition, this option is only suitable when hydrogen egress during the resting time is negligible. 

% *****************************************************************
\section{Results: Validation}
\label{Sec:Results1}
% *****************************************************************

This section presents a thorough validation analysis of the outputs of TDS Simulator. This verification exercise addresses, step-by-step, each of the elements of the formulation implemented. In Section \ref{Sec:Lattice}, the lattice diffusion model without traps is validated against analytical results, first presented by Kirchheim \cite{kirchheim2016bulk}. Then, the predictions obtained with Oriani's model for a one-trap system are benchmarked against the numerical results from Raina et al. \cite{Raina2018} in Section \ref{Sec:Oriani1trap}. Subsequently, in Section \ref{Sec:McNabb-Foster}, the McNabb and Foster implementation is validated against the results obtained by Legrand et al. \cite{Legrand2015} for a one-trap system. Multi-trap predictions based on Oriani's model are validated, in Section \ref{Sec:OrianiMultiTrap}, against results obtained by Drexel and co-workers using finite differences \cite{drexler2021critical}. The literature is scarce on the analysis of multiple-trap systems using McNabb-Foster \cite{Turnbull1997,ebihara2014numerical}. While this strengthens the novelty of this work, it hinders verification. To this end, we conduct in Section \ref{Sec:MultiTrap} an analysis of a two-trap alloy with both Oriani and McNabb-Foster models, showing how the predictions of the latter converge to the former for a sufficiently high value of $\nu$.

% -.-.-.-.-.-.-.-.-.-.-.-.-.-.-.-.-.-.-..-.-.-.-.-.-.-.-.-.-.-.-.-.-.-.-.-.-.-
\subsection{Lattice desorption}
\label{Sec:Lattice}
% -.-.-.-.-.-.-.-.-.-.-.-.-.-.-.-.-.-.-..-.-.-.-.-.-.-.-.-.-.-.-.-.-.-.-.-.-.-

\begin{figure}[H]
	\centering
	\includegraphics[scale=1.1]{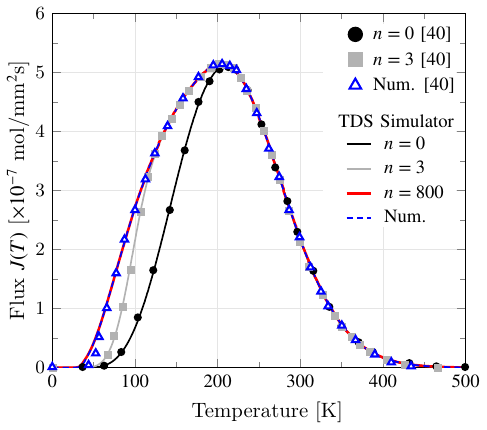}
	\caption{TDS desorption spectrum predictions for the case of an alloy without hydrogen traps. Comparison of present results (lines) with Kirchheim~\cite{kirchheim2016bulk} (symbols). The perfect agreement obtained validates the implementation of the lattice (no traps) model.}
	\label{Results_no_traps}	
\end{figure}

The implementation of the analytical formulation previously introduced in Section~\ref{Notraps} is validated with the results reported by Kirchheim~\cite{kirchheim2016bulk} in Fig.~\ref{Results_no_traps}. Following Ref. \cite{kirchheim2016bulk}, the simulation parameters comprise $L = 100$ mm, $\phi = 0.001$ K/s, $C_L^0 = 0.001$ mol/mm$^3$, $D_0 = 0.5$ mm$^2$/s, and $E_L=4150$ J/mol. As in Ref. \cite{kirchheim2016bulk}, we produce analytical estimations considering one ($n$ = 0) and three ($n$ = 3) terms. Additionally, an approximation of the case $n \rightarrow \infty$ considering $n=800$ terms as well as a numerical result obtained by solving the McNabb-Foster model in Section~\ref{Sec_McNabb_Foster} with no sink term are included. As shown in Fig. \ref{Results_no_traps}, excellent agreement is attained with the work by Kirchheim~\cite{kirchheim2016bulk}, verifying the lattice (no traps) implementation. Also, as reported by Kirchheim~\cite{kirchheim2016bulk}, it is evident that the agreement between the numerical solution and the analytical one improves as the number of terms considered in Eq. (\ref{notrap1}) increases. It is worth noting that this lattice-only curve cannot be captured by a Gaussian function, highlighting the need for more suitable approaches to infer lattice and individual trap type contributions (a typical Gaussian-based approach would have wrongly inferred trapping contributions from the `shoulder' on the left side of the figure). 

% -.-.-.-.-.-.-.-.-.-.-.-.-.-.-.-.-.-.-..-.-.-.-.-.-.-.-.-.-.-.-.-.-.-.-.-.-.-
\subsection{Oriani; single trap system}
\label{Sec:Oriani1trap}
% -.-.-.-.-.-.-.-.-.-.-.-.-.-.-.-.-.-.-..-.-.-.-.-.-.-.-.-.-.-.-.-.-.-.-.-.-.-

\begin{figure}[H]
	\centering
	\includegraphics[scale=1.1]{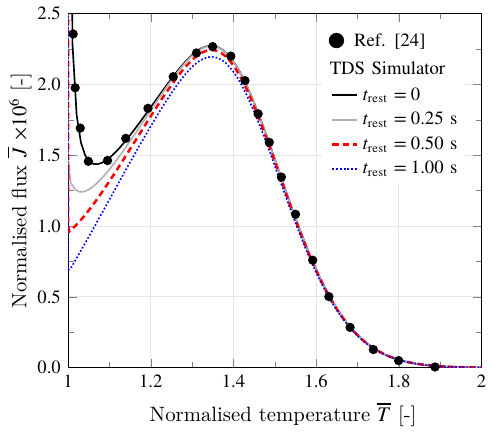}
	\caption{TDS simulation with varying rest period $t_{\text{rest}}$. The non-dimensional flux $\overline{J}$ is plotted against the non-dimensional temperature $\overline{T}$ (refer to \ref{App:NonDimensional} for non-dimensional quantities). The analysis also serves to validate the Oriani implementation, as data for $t_{\text{rest}}=0$ has been reported by Raina \textit{et al}.~\cite{Raina2018} (Fig. S2 in their Supplementary Material).}
	\label{Results_1_trap_Raina}	
\end{figure}

The implementation of Oriani's model, presented in Section~\ref{Oriani}, is validated against the results reported by Raina \textit{et al}.~\cite{Raina2018} for one hydrogen trap ($n_t=1$). Since Oriani's model was formulated in non-dimensional form in Ref. \cite{Raina2018}, the results are compared in Fig.~\ref{Results_1_trap_Raina} in terms of non-dimensional TDS flux versus temperature curves (see~\ref{App:NonDimensional} for a complete description of the non-dimensional formulation). The parameters used in the simulations, taken from Ref. \cite{Raina2018}, are $\overline{\Delta H}=-10$, $\theta_L^0=10^{-6}$, $\overline{E}_L=2.75$, $\overline{\phi}=0.1$, and $\overline{N}=10^{-3}$. Additionally, to assess the effect of the resting time defined in Eq.~(\ref{trest}), the flux estimates for four different magnitudes of the resting time ($t_{\text{rest}}$) have been included in Fig.~\ref{Results_1_trap_Raina}. An excellent fit is attained between the predictions by TDS Simulator and the results reported by Raina \textit{et al}.~\cite{Raina2018}, both for $t_{\text{rest}}=0$. Note that, when the resting time is not considered, the flux curve exhibits an initial spike due to the rapid desorption of lattice hydrogen. This spike attenuates as the resting time increases, bringing a concomitant decrease in the initial flux. The present case study validates the implementation of the Oriani transport model.

% -.-.-.-.-.-.-.-.-.-.-.-.-.-.-.-.-.-.-..-.-.-.-.-.-.-.-.-.-.-.-.-.-.-.-.-.-.-
\subsection{McNabb-Foster; single trap system}
\label{Sec:McNabb-Foster}
% -.-.-.-.-.-.-.-.-.-.-.-.-.-.-.-.-.-.-..-.-.-.-.-.-.-.-.-.-.-.-.-.-.-.-.-.-.-

\begin{figure}[H]
	\centering
	\includegraphics[scale=1.1]{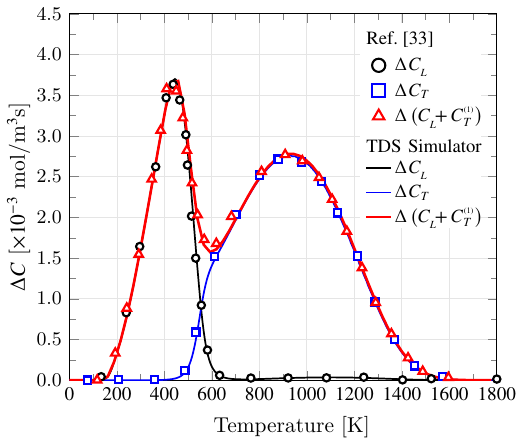}
	\caption{Validation of the McNabb-Foster implementation. Comparison between TDS desorption spectrum predictions for the lattice and trapped hydrogen reported by Legrand \textit{et al.}~\cite{Legrand2015} (symbols) and those obtained with TDS Simulator (lines).}
	\label{Results_1_trap_McNabb_Foster}	
\end{figure}

We proceed to validate the implementation of the McNabb-Foster diffusion model in Section~\ref{Sec_McNabb_Foster} against the results reported by Legrand \textit{et al.}~\cite{Legrand2015} for one single hydrogen trap. In this case, the input quantities, as per Ref.~\cite{Legrand2015}, are $D_0=2.74 \times 10^{-6}$ m$^2$/s, $N_L=1.27\times 10^{29}$ sites/m$^3$, $N_T=1.2\times 10^{24}$ sites/m$^3$, $L=4$ mm, $\phi=50$ K/min, $E_t=E_L=19.29$ kJ/mol, $E_d=53.69$ kJ/mol ($\Delta H=-44.4$ kJ/mol), $C_L^0=1$ mol/m$^3$, $\nu_t = 0.1$ GHz, and $\nu_d = 10$ THz. In this case, we do not use Oriani's equilibrium to define the initial trap occupancy, as implemented by default in TDS Simulator, and instead set it to $\theta_{T,0}^{(1)}=1$, as in Ref. \cite{CS2020b}. The obtained results are shown in Fig.~\ref{Results_1_trap_McNabb_Foster} in terms of the quantity of hydrogen that escaped the simulated TDS specimen at each time (desorption rate), as computed using Eq. (\ref{desprate}). The results show an excellent agreement with those reported by Legrand \textit{et al.}~\cite{Legrand2015}, verifying the implementation of the McNabb-Foster formulation.

\subsection{Oriani; multi-trap system}
\label{Sec:OrianiMultiTrap}

\begin{figure}[H]
	\centering
	\includegraphics[scale=1.1]{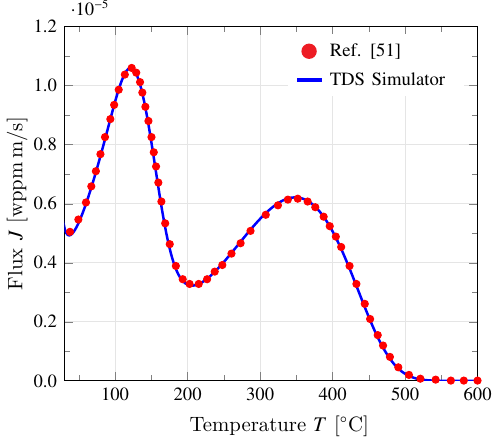}
	\caption{Validation of the Oriani implementation for multi-trap systems. Comparison between the hydrogen flux predictions reported by Drexler \textit{et al.}~\cite{drexler2021critical} (markers) and those obtained with TDS Simulator (solid line).}
	\label{Results_2_traps_drexler}	
\end{figure}

To validate the implementation of the Oriani model for multi-trap systems, the predictions by TDS Simulator are benchmarked against the results obtained by Drexler and co-workers using the finite differences method \cite{drexler2021critical}. Mimicking Fig. 6d in Ref. \cite{drexler2021critical}, we consider a two-trap system with the following parameters: $D_0=0.133\times 10^{-6}$ m$^2$/s, $L=1$ mm, $E_t^{(1)}=E_t^{(2)}=E_L=5.63$ kJ/mol, $\phi=2$ K/s, $N_L=1.2291\times 10^{29}$ atom/m$^3$, $N_T^{(1)}=6.0221\times 10^{25}$ sites/m$^3$, $\Delta H^{(1)}=-30$ kJ/mol ($E_d^{(1)}=35.63$ kJ/mol), $N_T^{(2)}=6.0221\times 10^{24}$ sites/m$^3$, $\Delta H^{(2)}=-70$ kJ/mol ($E_d^{(2)}=75.63$ kJ/mol), and $C_L^0=0.1$ mol/m$^3$. The output of this validation exercise is provided in Fig.~\ref{Results_2_traps_drexler}, with the blue curve denoting the results obtained with TDS Simulator and the red circles being the output of the finite difference calculations by Drexler et al. \cite{drexler2021critical}. A very good fit is found between the reference results and the predictions by TDS Simulator, demonstrating the validity of the implemented formulation.

\subsection{McNabb-Foster; multi-trap system}
\label{Sec:MultiTrap}

\begin{figure}[H]
	\centering
	\includegraphics[scale=1.1]{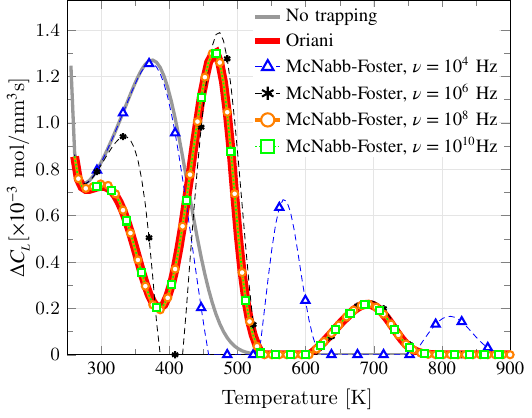}
	\caption{Comparison of TDS simulations considering no trapping, Oriani's model, and the McNabb-Foster formulation for different values of jump frequency $\nu_t=\nu_d=\nu$. The predictions obtained with the McNabb-Foster model for frequencies of $\nu=10^8$ Hz (orange circles) and $\nu=10^{10}$ (green squares) are in perfect agreement with the predictions obtained upon assuming Oriani's equilibrium.}
	\label{Results_2_trap_McNabb_Foster}	
\end{figure}

Finally, to verify our implementation of the McNabb-Foster model for multi-trap systems, we run simulations for the same multi-trap system using both Oriani and McNabb-Foster models to see if the latter converges to the former when $\nu_d,\nu_t \rightarrow \infty$, as expected theoretically. To this end, an alloy with two trap types and the following parameters is considered: $D_0=2.74\times 10^{-6}$ m$^2$/s, $L=4$ mm, $E_t^{(1)}=E_t^{(2)}=E_L=19.29$ kJ/mol, $\phi=0.2$ K/s, $N_L=1.27\times 10^{29}$ atom/m$^3$, $N_T^{(1)}=1.2\times 10^{24}$ sites/m$^3$, $\Delta H^{(1)}=-44.4$ kJ/mol ($E_d^{(1)}=63.69$ kJ/mol), $N_T^{(2)}=2.2\times 10^{24}$ sites/m$^3$, $\Delta H^{(2)}=-74.4$ kJ/mol ($E_d^{(2)}=93.69$ kJ/mol), $C_L^0=1$ mol/m$^3$, and $\theta_T^0=1$. Then, for the McNabb-Foster model, the jump frequency is taken to be the same for the two trap types and for trapping and detrapping (i.e., $\nu=\nu_d^{(1,2)}=\nu_t^{(1,2)}$), and is varied from $10^4$ to $10^{10}$ Hz. The results obtained, shown in Fig. \ref{Results_2_trap_McNabb_Foster}, reveal clear agreement between Oriani and McNabb-Foster predictions for sufficiently high values of the jump frequency $\nu$. Specifically, the simulations conducted here show that a jump frequency equal to or larger than $\nu=10^8$ Hz is sufficiently large for trap kinetics to occur on a much smaller time scale than hydrogen diffusion, with the $\nu=10^8$ (orange circles) and $\nu=10^{10}$ (green squares) results overlapping with the prediction based on Oriani's equilibrium (red line). It should be emphasized that this value of $\nu=10^8$ Hz is much smaller than the Debye frequency ($\nu=10^{13}$ Hz), suggesting that Oriani's equilibrium generally holds. This is in agreement with the findings by Toribio and Kharin \cite{toribio2015generalised}, who concluded in their generalised trapping analysis that Oriani's model is meaningful for bcc steels.

% *****************************************************************
\section{Results: experimental calibration and usage examples}\label{Sec:Results2}
% *****************************************************************

We now showcase how TDS Simulator can be employed to determine trapping characteristics from experimental TDS spectra by analysing two datasets corresponding to two martensitic steels, 4340 high-strength tempered martensitic steel (Section \ref{Sec:Experiments1}) and a martensitic steel containing Ti carbides (Section \ref{Sec:Experiments2}). 

\subsection{Trapping characteristics of a 4340 tempered martensitic steel}
\label{Sec:Experiments1}

For the first example, TDS Simulator is used to analyse the TDS data obtained by Novak et al. \cite{Novak2010} for a high-strength tempered 4340 steel. The first step is to digitise the experimental $\Delta C$ (wt ppm/min) vs $T$ (C) data, convert it to wt ppm/s vs C, and store it in a 2-column \texttt{.txt} file. The user must then open TDS Simulator and make appropriate choices. This means, in the \texttt{Simulation} tab, to select the appropriate graphical output ($\Delta C$, in this case, with the default $C_L+C_T$ option), and $x$-axis quantity (temperature). The units must then be selected as wppm/s and Celsius. One then goes to the \texttt{Fit experimental data} tab, makes appropriate choices for the second column variable (in this case: $\Delta C$, \texttt{DeltaC}) and the units (C and wppm/s for the first and the second column, respectively), and clicks \texttt{Load experimental data} to select the \texttt{.txt} file. The experimental data will be automatically plotted on the right side of the GUI.\\

The next step is to define the analysis parameters in the \texttt{Model parameters} tab. The thickness of the sample was reported to be $L=0.0063$ m \cite{Novak2010}. As shown in Appendix A of Ref. \cite{Novak2010}, multiple heating rates were considered; here, we focus first on the results for the 200 $^{\circ}$C/h ($\phi=0.055$ K/s) case, as it provides the richest TDS output, and then consider the other two heating rates (100 and 50 $^{\circ}$C/h). The resting time is not provided so a typical value of 45 min. is considered ($t_{\text{rest}}=2700$ s). The experimental data goes from a minimum temperature of approximately 20 $^{\circ}$C to a maximum temperature of 620 $^{\circ}$C; thus, we define $T_{\text{min}}=293$ K and $T_{\text{max}}=893$ K. The default numerical inputs are assumed; i.e., a number of temperature evaluations of 200 and a 100-element discretisation. In terms of material inputs, the properties characteristic of the ferritic lattice are appropriate for martensitic steels, and even more for tempered martensitic steels, as is the case here. Therefore, the values adopted are those listed for the bcc iron material family in Table \ref{Tab:properties}; that is, activation energy for lattice diffusion $E_L=5690$ J/mol, pre-exponential diffusion factor $D_0=7.23\times 10^{-8}$ m$^2$/s, molar mass $M_M=55.847$ g/mol, mass density $\rho_M=7.8474$ g/cm$^3$, and lattice site density $N_L=5.1\times 10^{29}$ atom/m$^3$. No details are provided of the charging condition, and therefore the value for the initial lattice hydrogen concentration listed in Table \ref{Tab:properties} for bcc iron is taken as a good approximation, $C_L^0=0.06$ mol/m$^3$; other choices of $C_L^0$ are also considered later on to assess the importance of this assumption. The number of traps is taken to be six; as discussed later, the fitting algorithm will automatically identify how many traps are needed to reproduce the TDS spectrum and therefore picking the highest number (6) is advisable. Finally, it remains to select the hydrogen transport model and define the characteristics of each trap type. Regarding the former, we choose to select Oriani in the \texttt{Simulation} tab, based on the analysis of Section \ref{Sec:MultiTrap}, which shows that Oriani's assumption of equilibrium is generally reasonable. Regarding the trapping variables, we will make use of the fitting capabilities of TDS Simulator in this example and therefore leave unchanged the default values (a trap binding energy of $\Delta H=-54.3$ kJ/mol and a trap density of $N_T=1.5 \times 10^{25}$ sites/m$^3$, for all trap types).\\

The options available for the fitting procedure are provided in the \texttt{Fit experimental data} tab. We choose to adopt the default options: the global optimization algorithm, with a maximum number of iterations equal to 150, a population size of 400, and a tolerance of $10^{-11}$. As per the guidelines above (see Section \ref{Sec:OptimizationProc}), the \texttt{Update initial H concentration} box is left unticked. Upon clicking the \texttt{Fit model} button, TDS Simulator displays a `Please wait. Fitting models...' message, and information about the fitting procedure is provided in MATLAB's \texttt{Command Window} and through the new window that TDS Simulator opens, as shown on the right side of Fig. \ref{Software_graph_2}. Specifically, MATLAB's \texttt{Command Window} displays statistics describing the calculations in each iteration, including \texttt{Iteration} (Iteration number), \texttt{f-count} (cumulative number of objective function evaluations), \texttt{Best f(x)} (best objective function value), \texttt{Mean f(x)} (mean objective function value over all particles), and \texttt{Stall Iterations} (number of iterations since the last change in \texttt{Best f(x)}). The new window also displays the experimental curve (using a red, dashed line) and the best TDS curves obtained for each iteration by the optimisation algorithm (dark blue line for the last best solution, and light blue lines for the solutions in previous iterations), allowing the user to readily visualise how the simulated curve approaches the experimental one.\\

\begin{figure}[H]
	\centering
	\includegraphics[scale=1.0]{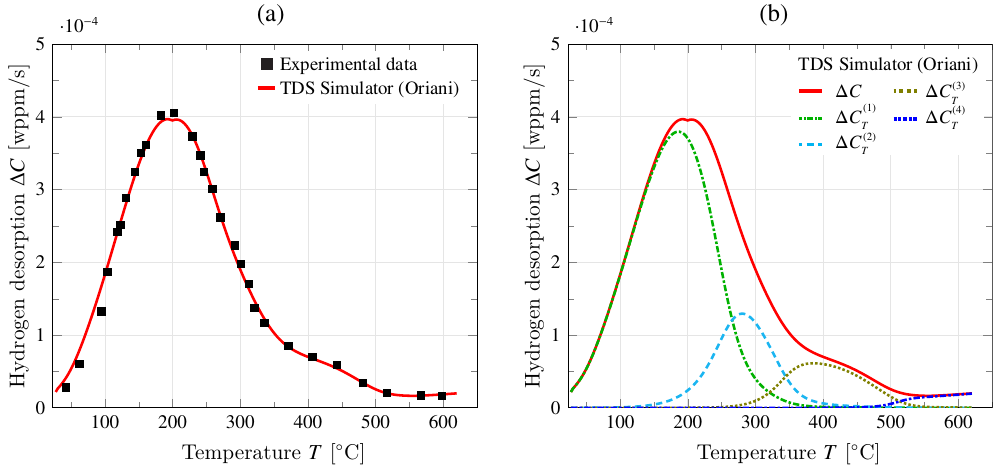}
	\caption{Using the inference fitting capabilities of TDS Simulator to gain insight into the trapping characteristics of a martensitic steel: (a) experimental \cite{Novak2010} and simulated desorption curves, with the latter being obtained using TDS Simulator's optimization algorithm, and (b) contribution of each relevant type, as determined by TDS Simulator. In (b), traps are ordered from smaller to larger absolute binding energy $|\Delta H|$.}
	\label{Experimental_Novak1}	
\end{figure}

The result of the fitting algorithm is shown in Fig. \ref{Experimental_Novak1} (a), together with the experimental curve from Novak et al. \cite{Novak2010}. The optimisation algorithm ends after 150 iterations, with a computation time in the order of 45 minutes on a standard desktop computer. The results presented in Fig. \ref{Experimental_Novak1} (a) show that the TDS Simulator fitting algorithm delivers a very good agreement with the experiment, accurately capturing all the peaks observed in the TDS spectrum, from the largest one at approximately 200 $^{\circ}$C to the smallest ones that appear at higher temperatures. By clicking on the \texttt{Hydrogen traps} tab, the user can see what trap binding energies ($\Delta H$) and densities ($N_T$) have been found by the algorithm to better describe the TDS curve. In this case, it can be seen that the algorithm assigns a very small binding energy ($|\Delta H| \sim$ 19 kJ/mol) to two trap types, and upon plotting their contributions to the desorption curve (\texttt{Simulation} tab, \texttt{CT} graphical output), one can see that these are negligible or non-existent, with the TDS data being governed by the remaining four other trap types. This is because their peaks appear at lower temperatures, below those considered in the test, as it can be captured by extending the temperature range in the simulation. Hence, TDS Simulator can be used to determine the number of relevant trap types needed to rationalise a given TDS curve. This can be readily seen in Fig. \ref{Experimental_Novak1} (b), where the contributions of the relevant trap types are shown superimposed. These dominant trap types are described by the following trapping parameters: $\Delta H^{(1)}=-53.1$ kJ/mol, $N_T^{(1)}=5.19 \times 10^{24}$ sites/m$^3$, $\Delta H^{(2)}=-68.7$ kJ/mol, $N_T^{(2)}=1.23 \times 10^{24}$ sites/m$^3$, $\Delta H^{(3)}=-91.7$ kJ/mol, $N_T^{(3)}=7.72 \times 10^{23}$ sites/m$^3$, $\Delta H^{(4)}=-140.1$ kJ/mol, $N_T^{(4)}=5.12 \times 10^{23}$ sites/m$^3$. It should be noted that the algorithm does not provide traps in any particular order; for the sake of clarity, we have followed the numbering of Fig. \ref{Experimental_Novak1} (b), where traps are ordered from smaller to larger absolute binding energy $|\Delta H|$.\\

\begin{figure}[htp]
	\centering
	\includegraphics[scale=1.0]{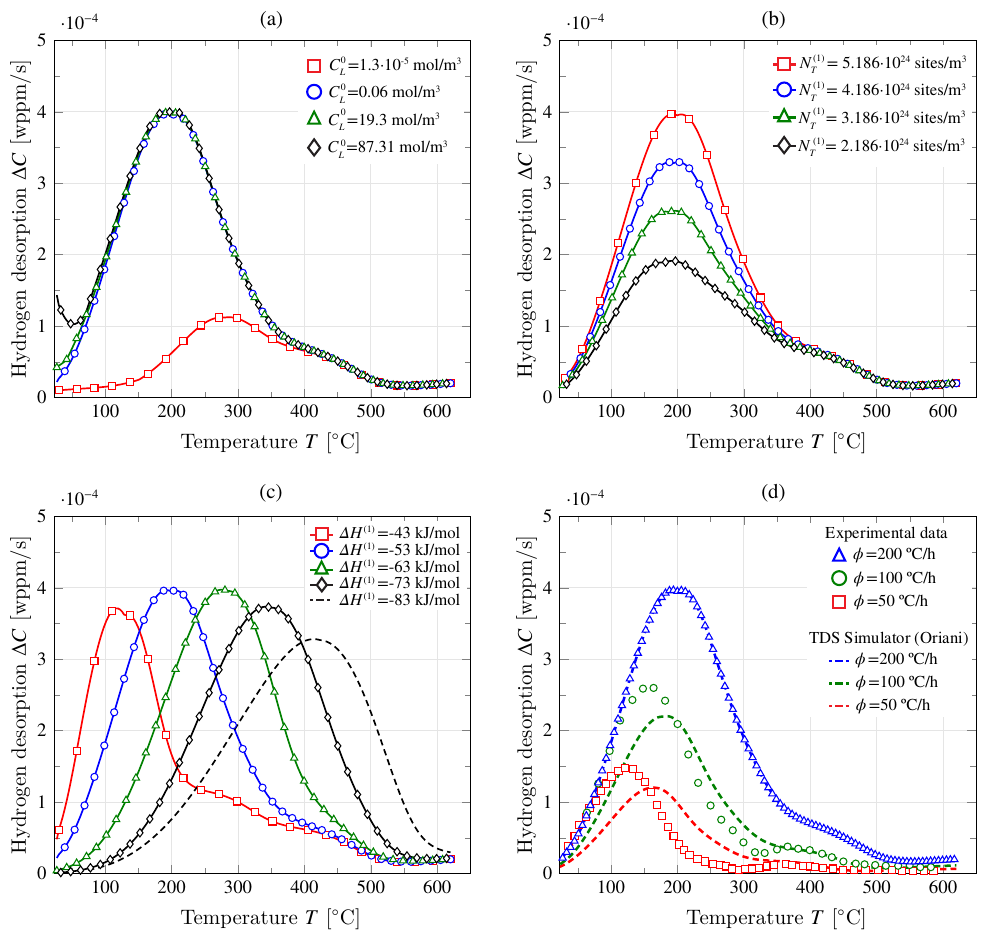}
	\caption{Using TDS Simulator to gain insight into the trapping characteristics of a martensitic steel: (a) role of the initial lattice hydrogen content $C_L^0$, (b) influence of the trap density $N_T$, (c) influence of the trap binding energy, and (d) comparison with experimental data \cite{Novak2010} at various heating rates.}
	\label{Experimental_NovakParametric}	
\end{figure}

The results shown in Fig. \ref{Experimental_Novak1} (b) illustrate how the first peak is mainly the effect of trap 1, the weakest trap among those observed in this temperature regime ($\Delta H^{(1)}=-53.1$ kJ/mol) but also the one with the highest trap density ($N_T^{(1)}=5.19 \times 10^{24}$ sites/m$^3$), which justifies the peak height. The second trap contributes to the widening of the peak observed at temperatures higher than 300 C. This second trap only spreads out the main peak and does not generate a new one because its binding energy ($\Delta H^{(2)}=-68.7$ kJ/mol) is very close to that of trap 1, and because of its notably smaller density ($N_T^{(2)}=1.23 \times 10^{24}$ sites/m$^3$). The TDS spectrum levels off slowly with increasing temperature due to the contribution of trap 3, which is of a higher absolute binding energy ($\Delta H^{(3)}=-91.7$ kJ/mol) but of a lower height, due to its lower density ($N_T^{(3)}=7.72 \times 10^{23}$ sites/m$^3$). Finally, there appears to be an uptick in hydrogen desorption rate at approximately 570 C, which is nicely captured by a fourth trap with the highest (absolute) binding energy ($\Delta H^{(4)}=-140.1$ kJ/mol) and the lowest density ($N_T^{(4)}=5.12 \times 10^{23}$ sites/m$^3$). Considering the underestimation in (absolute) trap binding energies associated with Kissinger's method, the values of $\Delta H$ attained are sensible values for high-strength, tempered martensitic steels \cite{Li2004}. With the strongest trap being typically identified as metal carbides, while traps 1 to 3 are within the range for martensitic interfaces, prior austenite grain boundaries, and mixed dislocation cores \cite{Li2004,Novak2010}. However, it should also be noted that the size of the last, stronger trapping contribution is very small, and could also be capturing an experimental artifact (for example, a non-subtracted and higher than usual baseline content).\\

The calibrated model is then used to conduct parametric studies that can provide further insight into TDS Simulator's predictions and the hydrogen trapping characteristics of high-strength martensitic steels. First, as shown in Fig. \ref{Experimental_NovakParametric} (a), the choice of initial lattice hydrogen concentration $C_L^0$ is assessed. The results demonstrate that predictions are largely insensitive to this choice for values spanning four orders of magnitude. When the magnitude is taken to be approximately 1500 times the one reported in Table \ref{Tab:properties} ($C_L^0=87.31$ mol/m$^3$, versus a reference value of $C_L^0=0.06$ mol/m$^3$), then a small deviation is observed at the beginning of the TDS experiment. Significant differences are only observed if the initial lattice content is taken to be approximately 4600 times smaller than the reference value ($C_L^0=1.3\times 10^{-5}$ mol/m$^3$). These results show that uncertainties in the quantification of $C_L^0$ have a negligible influence on predicted TDS results, and that the values provided in Table \ref{Tab:properties} constitute a reasonable starting point for TDS Simulator users. Then, in Fig. \ref{Experimental_NovakParametric} (b), the influence of the trap density is evaluated by varying the magnitude of $N_T^{(1)}$. A decrease in the hydrogen desorption rate is observed with decreasing trap density, in agreement with expectations. The role of the trap binding energy is evaluated in Fig. \ref{Experimental_NovakParametric}c, with results showing how increasing trap depth ($|\Delta H|$) shifts the desorption peak towards higher temperatures. Interestingly, some variations in peak height are also observed. Finally, the ability of the model to predict behaviour at various heating rates is investigated in Fig. \ref{Experimental_NovakParametric}d, as the reference experimental work by Novak et al. \cite{Novak2010} provides data also for 100 and 50 $^{\circ}$C/h. The simulation results capture the qualitative trend of the experiments but some quantitative differences are observed in terms of peak height and location. A better agreement can be achieved by extending the present \emph{mono-energetic} description to account for multi-energetic trapping energies - see Ref. \cite{kirchheim2016bulk}. The present modelling framework can also be enriched by considering interconnected trap systems with trap fluxes, as formulated by Toribio and Kharin \cite{toribio2015generalised}.

% ----------------------------------------------------------------------------
\subsection{The role of TiC trapping in martensitic steels}
\label{Sec:Experiments2}
% ----------------------------------------------------------------------------

The last case study deals with the fitting of a TDS curve obtained by Wei and Tsuzaki for a tempered martensitic steel containing titanium carbides (TiC) \cite{wei2012hydrogen,wei2004precise,wei2003hydrogen}. These precipitates are known to be strong, deep trapping sites, whose trapping capacity could be reflected by the presence of a second peak in the TDS spectra at high temperatures. The same protocol as in the previous experimental case study (Section \ref{Sec:Experiments1}) is followed. This involves first uploading the experimental TDS data (\texttt{Fit experimental data} tab), which is available as desorption rate ($\Delta C$, in wt ppm/s) versus temperature ($T$, in Celsius). The appropriate units and quantities are then selected in the \texttt{Simulation} tab, where Oriani's transport model is picked, as it has shown to generally hold (Section \ref{Sec:MultiTrap}). Then, we proceed to define the following test, numerical, and material inputs in the \texttt{Model parameters} tab. As per Refs. \cite{wei2012hydrogen,wei2004precise,wei2003hydrogen}, the sample thickness is $L=5$ mm and the tests are conducted using a heating rate of 100 $^{\circ}$C/h ($\phi=0.0278$ K/s). The resting time is approximately equal to $t_{\text{rest}}=120$ s, shorter than most TDS experiments, as Wei and Tsuzaki electroplated samples with cadmium to prevent hydrogen egress and the TDS experiment was run within a few minutes from the cadmium removal process. As shown below, this enables capturing the initial spike due to rapid lattice desorption discussed in Section \ref{Sec:Oriani1trap}. The initial and final temperatures are chosen to mimic the TDS experiment; i.e., $T_{\text{min}}=293$ K and $T_{\text{min}}=1100$ K \cite{wei2012hydrogen}. In terms of numerical inputs, the default values are adopted: 200 temperature evaluations and 100 elements. The material employed is a tempered martensitic steel. therefore the lattice-related material properties reported in Table \ref{Tab:properties} for bcc iron are suitable. This implies a lattice activation energy of $E_L=5690$ J/mol, a pre-exponential diffusion factor of $D_0=7.23\times 10^8$ m$^2$/s, molar mass $M_M=55.847$ g/mol, mass density $\rho_M=7.8474$ g/cm$^3$, and lattice site density $N_L=5.1\times 10^{29}$ atom/m$^3$. Finally, it remains to define the initial lattice hydrogen concentration $C_L^0$ and the number of trap types. Regarding the former, hydrogen was introduced into the samples using electrochemical charging, which hinders quantitative estimates. As in the previous case study, the value provided in Table \ref{Tab:properties} could be used as a first approximation. However, it is worth noting that in their experiments, Wei and Tsuzaki used a recombination poison (NH$_4$SCH) to enhance hydrogen uptake. Given the notable effect that NH$_4$SCH has in augmenting hydrogen ingress, a value of initial lattice content ten times the one provided in Table \ref{Tab:properties} is adopted; $C_L^0=0.6$ mol/m$^3$. The number of traps is chosen to be five, as four traps appear to be sufficient to describe the TDS spectrum of the martensitic steel studied in the previous section and the algorithm has the ability to automatically disregard the contribution from traps that are not required. No changes are made to the default trapping values since the fitting algorithm is used. The fitting procedure is used with the default parameters (150 iterations, a population size of 400 and a tolerance of 10$^{-11}$). The optimisation ends upon reaching 150 iterations, which takes slightly less than 40 min on a regular desktop computer.\\

\begin{figure}[H]
	\centering
	\includegraphics[scale=1.0]{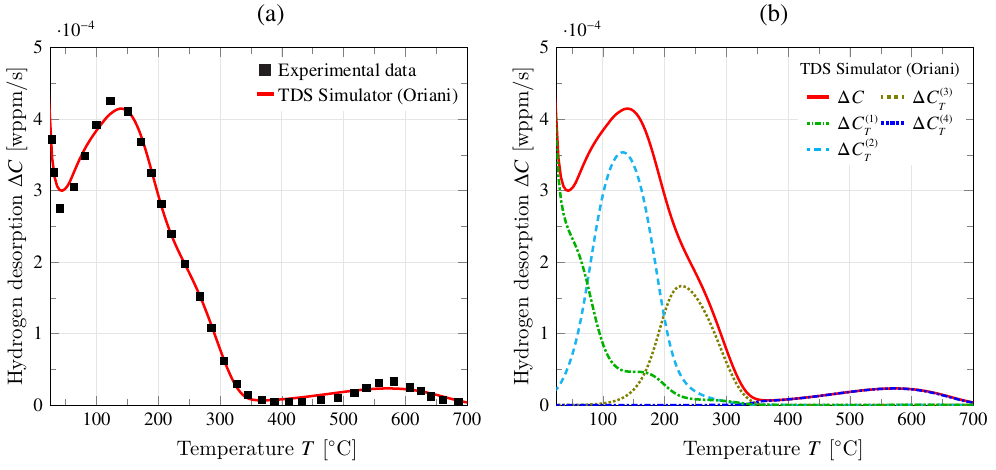}
	\caption{Using the inference fitting capabilities of TDS Simulator to gain insight into the trapping characteristics of a martensitic steel containing Ti carbides: (a) experimental \cite{wei2012hydrogen,wei2004precise,wei2003hydrogen} and simulated desorption curves, with the latter being obtained using TDS Simulator's optimization algorithm, and (b) contribution of each relevant type, as determined by TDS Simulator.}
	\label{Experimental_resultsJap}	
\end{figure}

The results of the analysis are given in Fig. \ref{Experimental_resultsJap}. First, Fig. \ref{Experimental_resultsJap} (a) shows the predicted TDS curve alongside the experimental data. As in the previous case study, TDS Simulator's fitting algorithm is shown to capture the characteristics of the experimental TDS spectrum very well, from the initial spike due to rapid desorption to the small peak arising as a result of trapping at TiC precipitates. A subsequent analysis of the trap parameters inferred by TDS Simulator reveals that the response can be captured by four traps, whose relevant contributions to the desorption curve are given in Fig. \ref{Experimental_resultsJap} (b). The results reveal that the sudden drop in desorption is governed by a shallow trap, with binding energy $\Delta H^{(1)}=-15$ kJ/mol and a very high density ($N_T^{(1)}=1.28 \times 10^{28}$ sites/m$^3$), in addition to the lattice hydrogen contribution. The TDS curve then experiences an uptick to develop its most prominent peak, which is mostly the consequence of the presence of a trap with a larger absolute binding energy ($\Delta H^{(2)}=-48.5$ kJ/mol) and a relatively large density ($N_T^{(2)}=7.19 \times 10^{24}$ sites/m$^3$). Similar to the previous case study, the TDS test data shows a smooth drop in the desorption rate with increasing temperature, due to the contribution of a nearby trap, in terms of binding energy ($\Delta H^{(3)}=-64.7$ kJ/mol), with a lower trap density $N_T^{(3)}=3.11 \times 10^{24}$ sites/m$^3$. The desorption rate becomes almost zero at approximately 400 $^{\circ}$C but then shows another peak at around 580 $^{\circ}$C, due to the presence of a deep trap with a strong binding energy ($\Delta H^{(4)}=-123.7$ kJ/mol) yet the smallest trap density ($N_T^{(4)}=8.85 \times 10^{23}$ sites/m$^3$). Once again, these values are consistent with the literature, considering the underestimation of $|\Delta H|$ values associated with Kissinger's method, with the strongest trap being incoherent TiC particles, and the remaining traps having values typical of grain boundaries, dislocations and coherent TiC particles. Since introducing strong traps into alloys is a strategy considered to design hydrogen-resistant materials \cite{chen2024hydrogen}, the combination of TDS experiments and the type of analysis provided by TDS Simulator can serve to accelerate the process. 

% *****************************************************************
\section{Conclusions}
\label{Sec:ConcludingRemarks}
% *****************************************************************
In this work, the first standalone software tool for simulating thermal desorption spectroscopy (TDS) experiments (TDS Simulator) is presented, which enables the quantification of hydrogen trapping characteristics in metals. TDS Simulator incorporates the two relevant hydrogen transport models, Oriani and McNabb-Foster, and can handle metallic systems with an arbitrary number of trap types. This also brings novelty on the theoretical side, as TDS analyses resolving trapping-detrapping kinetics (McNabb-Foster) have previously been limited to one trap type; despite all alloys having multiple trap types. TDS Simulator not only produces synthetic TDS data but also incorporates an inference algorithm to automatically determine trap binding energies and densities from existing experimental data. The toolbox is available as a MATLAB App, which includes a user-friendly GUI and can be freely downloaded by the community. As demonstrated in this work, the predictions of TDS Simulation agree with existing literature data, and all of the elements of the framework have been independently validated. The treatment of experimental data is also addressed, to showcase the usage of the software and highlight its strengths in providing the first toolbox for the automatic quantification of trapping characteristics. In addition, its unique capabilities (e.g., multiple theories and trap sites, ability to predict behaviour over a wider range of temperatures) are exploited to gain new fundamental insights. Among others, the results obtained show that,
\begin{itemize}
    \item Oriani and McNabb-Foster give identical results for sensible values of the jump frequency, implying that trap kinetics occur on a much smaller time scale than diffusion. 
    \item The rapid desorption drop typically predicted in numerical TDS analyses is not observed in experiments due to the resting time.
    \item The use of hydrogen transport models and a suitable inference algorithm enables quantifying trapping data (binding energies $\Delta H$ and densities $N_T$) from a single TDS experiment.     
    \item TDS spectra and associated trapping characteristics ($\Delta H$, $N_T$) are relatively insensitive to the initial lattice hydrogen concentration, reducing the sensitivity to hydrogen charging conditions.
    \item Desorption peak heights and locations are respectively governed by the trap density and binding energy, with the latter also having an influence on the desorption rate magnitude.
    \item Through its optimisation module, TDS Simulator can automatically determine how many trap types are needed to describe and rationalise a given TDS curve.\\

\end{itemize}

The present work describes a new software tool that will enable a better understanding of hydrogen-material interactions, which is of relevance to the development of hydrogen-compatible materials for the energy transition and the prediction of hydrogen-assisted failures, which is a pervasive problem across sectors, including defence, transport, nuclear, and construction. TDS Simulator can be freely downloaded at \url{https://mechmat.web.ox.ac.uk/codes}.
% *****************************************************************
\section*{Acknowledgments}\label{Sec:Acknowledgeoffunding}
% *****************************************************************
E. Mart\'{\i}nez-Pa\~neda would like to acknowledge helpful discussions with Dr Andres D\'{\i}az (University of Burgos), Dr Chuanjie Cui (University of Oxford), and Prof Kaneaki Tsuzaki (NIMS) and Dr Fugao Wei-San (Nippon Yakin Kogyo) for kindly providing additional information about their TDS experiments. The authors acknowledge financial support from the EPSRC (grants EP/V04902X/1 and EP/V009680/1). E. Mart\'{\i}nez-Pa\~neda additionally acknowledges financial support from UKRI’s Future Leaders Fellowship programme [grant MR/V024124/1] and from the UKRI Horizon Europe Guarantee programme (ERC Starting Grant \textit{ResistHfracture}, EP/Y037219/1). 

% ----------------------------------------------------------------------------
% ----------------------------------------------------------------------------
% ----------------------------------------------------------------------------
% ----------------------------------------------------------------------------
% ----------------------------------------------------------------------------
% ----------------------------------------------------------------------------
\appendix

\section{Non-dimensional form of the governing equations} \label{App:NonDimensional}

The previously introduced general multi-trap diffusion formulations can be written in non-dimensional form considering the following non-dimensional parameters:
\begin{equation}
    \overline{t} = \frac{T}{T_0} \, ; \,\,\,\,\,\,\,\,
    \overline{\phi}=\frac{\phi L^2}{T_0 D_0} \,\,\,\,\,\,\,\,
    \overline{\theta_L} = \frac{\theta_L}{\theta_L^0} \, ; \,\,\,\,\,\,\,\,
    \overline{\theta_T^{(i)}} = \frac{\theta_T^{(i)}}{(\theta_T^0)^{(i)}} \, ; \,\,\,\,\,\,\,\,
    \overline{E_L}=\frac{E_L}{RT_0}  \, ;\,\,\,\,\,\,\,\,
    \overline{D_L} = \frac{D_L}{D_0}
\end{equation}
\begin{equation}
    \overline{x} = \frac{x}{L} \, ; \,\,\,\,\,\,\,\, \overline{t} = \frac{t D_0}{L^2} \, ; \,\,\,\,\,\,\,\,
\overline{\Delta H} = \frac{\Delta H}{R T_0} \, ; \,\,\,\,\,\,\,\,
\overline{N_T^{(i)}} = \frac{N_T^{(i)}}{N_L} \, ; \,\,\,\,\,\,\,\,
\overline{\nu}_{t,d}^{(i)} = \frac{\nu_{t,d}^{(i)}L^2}{D_0}
\end{equation}

On this basis, the extended diffusion equation (\ref{Modif_Diff}) can be rewritten as:
\begin{equation}\label{Modif_Diff_nondim}
  \frac{\partial \overline{\theta_L}}{\partial \overline{t}} +  \sum_{i=1}^{n_t} \overline{N_T^{(i)}}\frac{\theta_T^0}{\theta_L^0}\frac{\partial \overline{\theta_T^{(i)}}}{\partial \overline{t}}   = \overline{D_L} \frac{\partial^2 \overline{\theta_L}}{\partial \overline{x}^2},
\end{equation}

\noindent and Oriani's equilibrium equation, Eq.  (\ref{Oriani_a}), as:
\begin{equation}\label{Oriani_a_nondim}
\frac{(\theta_T^0)^{(i)}\overline{\theta_T^{(i)}}}{1-(\theta_T^0)^{(i)}\overline{\theta_T^{(i)}}} = \frac{\theta_L^0\overline{\theta_L}}{1-\theta_L^0\overline{\theta_L}}\overline{K_T^{(i)}},
\end{equation}

\noindent with $K_T^{(i)}=\overline{K_T^{(i)}} = \exp \left(-\overline{\Delta H^{(i)}}/ \overline{T}\right)$. Inserting Eq.~(\ref{Oriani_a_nondim}) into Eq.~(\ref{Modif_Diff_nondim}), one can write the following non-dimensional PDE:
\begin{equation}\label{Oriani_c_nondim}
\frac{\partial \overline{\theta_L}}{\partial \overline{t}}\left\{1+\sum_{i=1}^{n_t} \frac{\overline{N_T^{(i)}} \, \overline{K_T^{(i)}}}{\left[1+\left(\overline{K_T^{(i)}}-1\right)\theta_L^0\overline{\theta_L} \right]^2}\right\}+\sum_{i=1}^{n_t}\frac{\overline{K_T^{(i)}} \, \overline{N_T^{(i)}} \overline{\Delta H^{(i)}} \overline{\phi}(\overline{\theta_L}-\theta_L^0\overline{\theta_L}^2)}{\overline{T}^2 \left[1+(\overline{K_T^{(i)}}-1)\theta_L^0\overline{\theta_L} \right]^2} = \overline{D_L}  \frac{\partial^2 \overline{\theta_L}}{\partial \overline{x}^2},
\end{equation}

Similarly, the McNabb-Foster's equilibrium equation, Eq. (\ref{eq:RainaMcNabbFoster_b}), can be rewritten as:
\begin{equation}\label{eq:RainaMcNabbFoster_b_nondim}
\frac{\partial \theta_T^0\overline{\theta_T^{(i)}}}{\partial \overline{t}}= \left[ \overline{k^{(i)}} \theta_L^0\overline{\theta_L} \left( 1 - (\theta_T^0)^{(i)}\overline{\theta_T^{(i)}} \right)  + \overline{p^{(i)}} (\theta_T^0)^{(i)}\overline{\theta_T^{(i)}}\left( 1 - \theta_L^0\overline{\theta_L} \right)  \right],
\end{equation}

\noindent with
\begin{equation}\label{eq:RainaMcNabbFoster_c_nondim}
\overline{k^{(i)}} = \overline{\nu_t^{(i)}} \exp \left( - \frac{E_t^{(i)}}{RT} \right), \quad \overline{p^{(i)}} = \overline{\nu_d^{(i)}} \exp \left( - \frac{E_d^{(i)}}{RT} \right).
\end{equation}

% ----------------------------------------------------------------------------
% ----------------------------------------------------------------------------

%% Text of bib
\end{document}